\documentclass{aa}

\usepackage{amsmath}
\usepackage{natbib}
\usepackage{graphicx}
\usepackage{txfonts}
\usepackage{xspace}
\usepackage{microtype}
\usepackage[colorlinks=true,linkcolor=red,urlcolor=black,citecolor=blue]{hyperref}

\bibpunct{(}{)}{;}{a}{}{,}

\newcommand{\gro}{GRO~J1008$-$57\xspace}
\newcommand{\suzaku}{\textsl{Suzaku}\xspace}
\newcommand{\nustar}{\textsl{NuSTAR}\xspace}
\newcommand{\swift}{\textsl{Swift}\xspace}
\newcommand{\rxte}{\textsl{RXTE}\xspace}
\newcommand{\msol}{\mathrm{M_\odot}}

\defcitealias{kuehnel2013a}{K13}
\defcitealias{kuehnel2014a}{K14}

\begin{document}

\title{Evidence for different accretion regimes in \gro}

\author{Matthias K\"uhnel \inst{1}
 \and Felix F\"urst \inst{2,3}
 \and Katja Pottschmidt \inst{4,5}
 \and Ingo Kreykenbohm \inst{1}
 \and Ralf Ballhausen \inst{1}
 \and Sebastian Falkner \inst{1}
 \and Richard E. Rothschild \inst{6}
 \and Dmitry Klochkov \inst{7}
 \and J\"orn Wilms \inst{1}
}

\authorrunning{K\"uhnel et al.}

\institute{
      Dr.\ Karl Remeis-Observatory \& ECAP, Universit\"at
      Erlangen-N\"urnberg, Sternwartstr. 7, 96049 Bamberg, Germany
 \and Cahill Center for Astronomy and Astrophysics, California
      Institute of Technology, Pasadena, CA 91125, USA
 \and European Space Astronomy Centre (ESA/ESAC), Operations Department,
      Villanueva de la Ca\~nada, 28692 Madrid, Spain
 \and CRESST/CSST/Department of Physics, UMBC, Baltimore, MD 21250, USA
 \and NASA Goddard Space Flight Center, Greenbelt, MD 20771, USA
 \and Center for Astrophysics and Space Sciences, University of
      California, San Diego, La Jolla, CA 92093, USA
 \and Institut f\"ur Astronomie und Astrophysik, Universit\"at
      T\"ubingen, Sand 1, 72076 T\"ubingen, Germany
}

\abstract{We present a comprehensive spectral analysis of the BeXRB
  \object{\gro} over a luminosity range of three orders of magnitude
  using \nustar, \suzaku and \rxte data. We find significant
  evolution of the spectral parameters with luminosity. In particular
  the photon index hardens with increasing luminosity at intermediate
  luminosities between $10^{36}$--$10^{37}$\,erg\,s$^{-1}$. This
  evolution is stable and repeatedly observed over different
  outbursts. However, at the extreme ends of the observed luminosity
  range, we find that the correlation breaks down, with a significance
  level of at least $3.7\sigma$. We conclude that these changes
  indicate transitions to different accretion regimes, which are
  characterized by different deceleration processes, such as Coulomb
  or radiation breaking. We compare our observed luminosity levels of
  these transitions to theoretical predications and discuss the
  variation of those theoretical luminosity values with fundamental
  neutron star parameters. Finally, we present detailed spectroscopy
  of the unique ``triple peaked'' outburst in 2014/15 which does not
  fit in the general parameter evolution with luminosity. The pulse
  profile on the other hand is consistent with what is expected at
  this luminosity level, arguing against a change in accretion
  geometry. In summary, \gro is an ideal target to study different
  accretion regimes due to the well constrained evolution of its
  broad-band spectral continuum over several orders of magnitude in
  luminosity.  }

\date{Received xx Month xxxx / Accepted xx Month xxxx}
\keywords{X-rays: binaries - stars: neutron - stars: Be - pulsars: individual: GRO J1008$-$57}

\maketitle

\section{Introduction}
\label{sec:intro}

The strong magnetic field of an accreting and pulsing neutron star
channels accreted matter onto its magnetic poles. There, infalling
plasma with velocities close to the speed of light is decelerated to
rest on the surface. The X-ray spectrum of accreting neutron stars
is therefore dominated by strong bulk-motion Comptonizaton caused by
the interaction of primary soft X-rays from the accretion mound or
column with hot electrons in the accretion column. A self-consistent
model of the emerging X-ray spectrum requires relativistic
magnetohydrodynamical calculations of this extreme plasma and is
still outside of the range of capabilities of current computing
facilities. However, in recent years some progress has been made and
models have become available in which the radiative transfer
equations are solved assuming thermal and bulk Comptonization
\citep{becker2007a,marcu2015a,wolff2016a} and which allow for
different (phenomenological) velocity profiles
\citep{farinelli2012a,farinelli2016a}. The details of the physical
mechanisms decelerating the accreted material to rest above the
neutron star's surface are the topic of current theoretical
investigations (see, e.g., discussions by \citealt{staubert2007a}
and \citealt{becker2012a}).

An ideal laboratory to investigate and check different theoretical
predictions about accretion physics is the class of transient Be X-ray
binaries (BeXRBs). Their luminosity changes by orders of magnitude
during an outburst and the spectral evolution of the neutron star in
different luminosity regimes can be studied in great detail. In
previous works, using all available \rxte data, we have shown that the
BeXRB \object{\gro} shows a tight correlation of its spectral shape
with the 15--50\,keV luminosity, which appears to be stable between
outbursts \citep[hereafter K13 and K14]{kuehnel2013a,kuehnel2014a}.
\gro was discovered with the \textsl{Compton Gamma Ray Observatory}
(\textsl{CGRO}) during a luminous outburst in 1993 July
\citep{wilson1994a,stollberg1993a}. The system consists of a neutron
star on a wide, eccentric orbit \citep[$a \sin i = 530\,$lt-s,
  $e=0.68$,][]{coe2007a} around a B0e type companion star
\citep{coe1994a}. In addition to the high eccentricity, the orbital
period of $P_\mathrm{orb} = 249.48$\,d \citepalias{kuehnel2013a} is
long enough for the circumstellar disk of the Be companion to be
tidally truncated at the 7:1 or 8:1 resonance \citep{okazaki2001a},
leading to regular type I outbursts. The source showed a type~II giant
outburst in 2012 November at an orbital phase of ${\sim}0.3$ with a
peak luminosity around 1\,Crab in the 15--50\,keV range
(Fig.~\ref{fig:batlc}).  After this outburst \gro showed three regular
type~I outbursts, the last of these on 2014 September \gro
\citep{nakajima2014a}. This expected outburst was followed by two
unusual type~II outbursts in 2014 November and 2015 January, occurring
within the same orbit \citep[][and
  Fig.~\ref{fig:batlc}]{nakajima2014b,kretschmar2015a}.  To our
knowledge, this ``triple-peaked'' outburst behavior had not been seen
in any other source, although some sources, such as A~0535+262
\citep[see, e.g.,][and references therein]{caballero2013a} or
GX~304$-$1 \citep{nakajima2012a,postnov2015a}, feature rare
``double-peaked'' outbursts. What causes the source to undergo
outbursts far away from periastron is not yet understood.  A
promising scenario proposed by \citet{okazaki2013a} is an inclined Be
disk with respect to the orbital plane.

In this work, we compare our previous findings to observations
performed by \suzaku, the Nuclear Spectroscopic Telescope Array
(\nustar), and \swift from 2012 to 2015 in order to discuss
implications on the theory of mass accretion onto magnetized neutron
stars. Sect.~\ref{sec:data} describes the data reduction of the
observations and energy selection criteria applied to the resulting
spectra. In addition, we discuss calibration uncertainties detected in
the spectra. The actual spectral analysis of \gro is presented in
Sect.~\ref{sec:spec} and the corresponding individual results and the
outburst behavior of the source are discussed in
Sect.~\ref{sec:discuss}. Section~\ref{sec:discuss:parevol} focuses on
the spectral evolution of \gro, where we compare the results of the
spectral analysis with our previous work and recent theoretical work.
We conclude the paper with the discovery of different accretion
regimes in this source.

\section{Observations \& Data Reduction}
\label{sec:data}

\begin{figure}
  \includegraphics[width=\columnwidth]{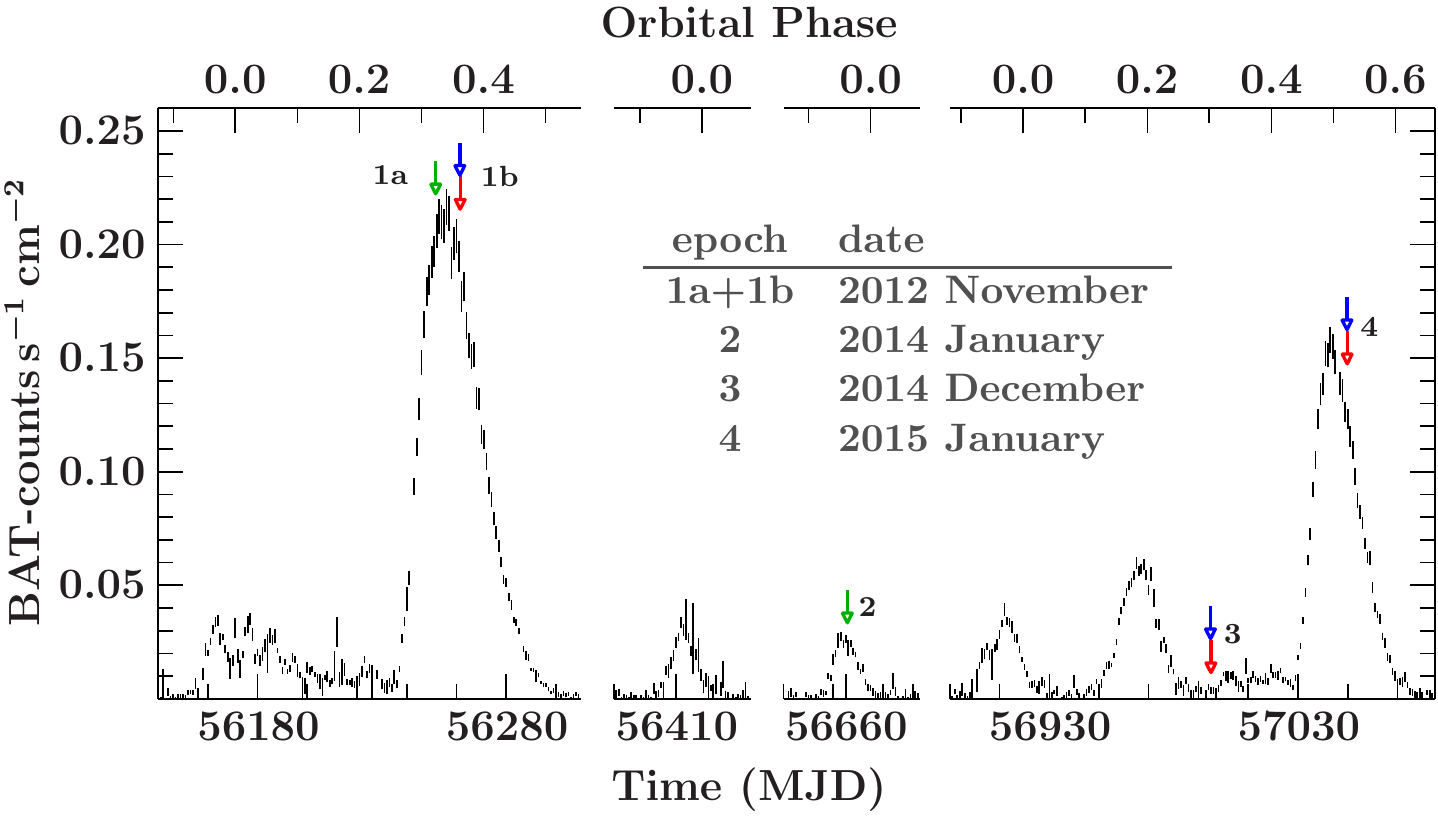}
  \caption{\swift-BAT \citep{krimm2013a} light curve showing the
    activity of \gro between 2012 August and 2015 February. The
    arrows on top mark the times of observations by \swift (red),
    \suzaku (green), and \nustar (blue). The numbers next to these
    arrows are the data epochs as defined in
    Table~\protect\ref{tab:obs}. The labeled dates correspond to
    these epochs.}
  \label{fig:batlc}
\end{figure}

Figure~\ref{fig:batlc} shows the light curves of \gro for all
outbursts between 2012 August and 2016 February as recorded by the
\swift Burst Alert Telescope (BAT). Arrows indicate the observations
studied here (see also Table~\ref{tab:obs}). During the peak of the
type~II giant outburst in 2012 November, \gro was observed with
\nustar, \suzaku, and \swift. After this type~II outburst, the
source went back to its normal behavior, featuring type~I outbursts
at every periastron passage. \suzaku observed the peak of such an
outburst in 2014 January. The next observation by \nustar, which was
simultaneous with \swift, was performed in 2014 December, after the
second outburst of the triple-peaked outburst. Finally,
\nustar and \swift also performed a joint observation slightly after
the peak of the third outburst, which reached around 700\,mCrab in
\swift-BAT during 2015 January.

The light curve of \gro during 2005, as measured by the Proportional
Counter Array \citep[PCA,][]{jahoda1996a} onboard \rxte (ObsID
90089-03-02-01) was used for the pulse profile analysis presented in
Sect.~\ref{sec:discuss:epochIV}. The GoodXenon events of the
Proportional Counter Units (PCU) 2 and 3 were accumulated into a light
curve with a time resolution of 1\,s. Finally, the light curve was
transformed into the barycenter of the solar system
\citepalias[see][for more details about the extraction of \rxte light
curves]{kuehnel2013a}.

\begin{table}
  \caption{List of observations used in this work. Data from epoch 1a has
    been used and analyzed by \citet{yamamoto2013a,yamamoto2014a} as well as by
    \citet{bellm2014a}, who used and analyzed epoch 1b in addition.}
  \label{tab:obs}
  \centering
  \begin{tabular}{lllll}
    \hline\hline
    Sat.\tablefootmark{a} & ObsID & Start (MJD)
      & Expt.\tablefootmark{b} & E\tablefootmark{c} \\
    \hline
    \suzaku & 907006010 & 56251.63 & 9061 & 1a \\
    \nustar & 80001001002 & 56261.36 & 14767 & 1b \\
    \swift & 00031030018 & 56261.53 & 1716 & 1b \\
    \suzaku & 408044010 & 56660.66 & 15322 & 2 \\
    \nustar & 90001003002 & 56994.82 & 26861 & 3 \\
    \swift  & 00081425001 & 56995.00 & 2263 & 3 \\
    \nustar & 90001003004 & 57049.74 & 17260 & 4 \\
    \swift & 00081425002 & 57049.79 & 1943 & 4 \\
    \hline
  \end{tabular}
  \tablefoot{
  \tablefoottext{a}{Name of the X-ray mission.}
  \tablefoottext{b}{Exposure time in seconds. For \nustar the longest FPM and
    for \suzaku the longest XIS exposure is given.}
  \tablefoottext{c}{Data epoch indicating which observations have been
    combined in the spectral analysis.}
  }
\end{table}

\subsection{\suzaku}
\label{sec:data:suzaku}

We reprocessed the \suzaku \citep{mitsuda2007a} data following the
Data Reduction (or ABC) Guide \citep{abcguide}. The data have been
calibrated using \texttt{aepipeline} as distributed by
\texttt{HEASOFT} v6.15.1. For the data of the \suzaku X-ray Imaging
Spectrometers \citep[XIS;][]{koyama2007a}, the XIS calibration
database (\texttt{CALDB}) released on 2014-12-22 was used. For the
calibration of the corresponding mirrors of the X-ray Telescopes
\citep[XRT;][]{serlemitsos2007a}, the releases 2011-05-05 (effective
area) and 2009-06-05 (Point Spread Function, PSF) are utilized. The
High X-ray Detector \citep[HXD;][]{takahashi2007a} data were
calibrated based on HXD CALDB release 2011-08-19.

Three of the four XIS (0, 1, and 3; data from XIS2 were no longer
available) were operated in the 1/4 window mode during the 2012
November and 2014 January observations of \gro. After extracting an
image for each XIS and all editing modes using \texttt{xselect}, we
have applied the attitude correction by \texttt{aeattcor2} using a
preliminary source region. The final source spectra of \gro were
extracted with \texttt{xselect} using a circular source region
centered at the source. The radii were set to $80''$ in XIS0 and
XIS3 for observation 408044010 and $90''$ otherwise. Pile-up needed
to be investigated for each XIS and editing mode separately and was
calculated for each each pixel using \texttt{pileest}. During
observation 408044010 pile-up of more than 4\% was avoided by annuli
source regions with inner radii of 13--$30''$. Due to the very high
count rate during observation 907006010, ellipsoidal areas within
the source region with more than 2\% pile-up were excluded. All
background regions were chosen to avoid source photons from the area
covered by the Point Spread Function (PSF) of the XRT with radii of
$90''$. With these regions as input for \texttt{xselect} the
XIS-spectra of the source and background were extracted. The
Redistribution Matrix Function (RMF) and Ancillary Response Function
(ARF), were generated with \texttt{xisrmfgen} and
\texttt{xissimarfgen} respectively. For the spectral analysis we
added the spectra of the different editing modes, ``$3\times3$'' and
``$5{\times}5$'', to generate one spectrum for each XIS.

The HXD \citep{kokubun2007a} consisted of two detectors, the PIN diodes and the
Gadolinium Silicate Crystals (GSO). Spectra for both instruments
were extracted using \texttt{hxdpinxbpi} and \texttt{hxdgsoxbpi},
respectively. These tools generate the background spectra based on
the Non X-ray Background (NXB), using modeled events of the
``tuned'' background v2.2 (PIN) and v2.6 (GSO), and Cosmic X-ray
Background (CXB), simulated following \citet{boldt1987a}. For PIN we
used epoch number 11 (2011-06-01) and for GSO epoch 2010-05-24 for
the response file (RSP). Additionally, the GSO ARF, calibrated on the
Crab pulsar from epoch 2010-05-26, was used.

We restricted the energy range of the \suzaku-observations
(epochs~1a and~2, see Table~\ref{tab:obs}) to 0.8--10\,keV for the
XISs, to 15--70\,keV for PIN, and to 60--100\,keV for GSO. The
XIS-spectra have been rebinned following \citet{nowak2011a}, i.e.,
each energy bin has at least a minimum signal-to-noise (SNR) of 8
and a minimum number of channels close to the half-width
half-maximum of the spectral resolution. For PIN and GSO, we applied
a channel binning resulting in a minimum SNR of 15 and 5 in each
energy bin, respectively.

During a preliminary spectral analysis we investigated possible
calibration uncertainties of the \suzaku-data. In both epochs,
residual features in all XISs are found around the Au-edge at
2.19--2.37\,keV, as already noticed by \citet{nowak2011a}.
Consequently, we ignored this energy range during the spectral
analysis. Further calibration uncertainties around the Si-edge at
1.72--1.88\,keV are known. Using the newer calibration, however,
this calibration feature is only present in the backside illuminated
XIS1, which we ignored appropriately in the analysis.
In epoch~2 (2014 January), the front illuminated XIS0 and 3 show 
residual emission at energies above 9\,keV in contrast to XIS1.
Thus, we ignored energies above 9\,keV in XIS0 and 3. In addition,
we found a slight discrepancy of the power-law photon index,
$\Gamma$, determined by the spectrum of the back illuminated XIS1 to
the other XISs in epoch~2. This behavior is already known \citep[see,
e.g., Sect.\ 5.2.1 of][]{tsujimoto2011a}, the photon index shift of
$\Delta \Gamma = 0.06^{+0.04}_{-0.07}$ is, however, only significant on the
${\sim}$90\% confidence level.

For epoch~1a we ignored PIN-data below 20\,keV because of a mismatch to
\nustar as seen by \citet{bellm2014a}, which is probably noise
contamination caused by the leakage current of the PIN sensors.
Furthermore, we found a broad-band wave-like structure in PIN with
an amplitude of {2--16\%} in flux,
which can be described by a Gaussian absorption centered at
$51.1^{+2.0}_{-5.8}$\,keV. \citet{kokubun2007a} noticed
this feature in data of the Crab pulsar (in epoch~1a, \gro's flux
was around 1\,Crab), which they attributed to an insufficient
modeling of the gadolinium fluorescent lines produced in the
detector. During the spectral analysis of \gro we modeled this
feature by the Gaussian as described above. In both epochs, the
GSO-data in the four energy bins between 75 and 81\,keV scatter
strongly around any predicted model flux. Because the GSO background
in this energy range is dominated by the decay of activated
$^{151}$Gd and $^{149}$Eu \citep{kokubun1999a} we ignored those bins
in our analysis.

Among the calibration features we have described above and taken
into account during the spectral fitting, we added 1\% systematic
uncertainties to the XIS-, PIN- and GSO-spectra to achieve a reduced
$\chi^2$ around unity. We have followed a slightly different
treatment of calibration uncertainties compared to
\citet{bellm2014a}. In particular, they have not taken the Gaussian
absorption feature caused by the gadolinium fluorescent lines and
the photon index shift in XIS1 into account, but added higher
systematic uncertainties of 3\% to the spectra. However, their fit
of the \suzaku-data during \gro's giant outburst (epoch~1a) still
resulted in a reduced $\chi^2$ of $3.15$ (see their Table~2).

\subsection{\nustar}
\label{sec:data:nustar}

Extraction of the \nustar data was performed separately for Focal
Plane Module (FPM) A and B, following the \nustar Data Analysis
Software Guide \citep{perri2015a}. The data were extracted using the
standard \texttt{nustardas} pipeline (v1.4.1) and CALBD 20150316 as
distributed with HEASOFT v6.16 and cleaned for source occultation
by the Earth and SAA passages. Spectra in mode 01 (SCIENCE) event
files were extracted for each of the three observations separately
using a region with 120$''$ diameter, centered on the J2000
coordinates of \gro. The background was extracted from a region of
the same size at the other end of the field of view, to avoid
contamination from source photons in the outer wings of the PSF. We
additionally extracted data from mode 06 (SCIENCE\_SC), in which the
optical bench star tracker is occulted, and therefore the sky image
reconstruction is solely based on the spacecraft attitude. This
results in a smeared point source in the sky image. However, as
the metrology system is still functional, the effective area is
calculated correctly \citep[see][for details about mode 06 extract]{walton2016a}.
We used a source region with 150$''$ diameter
to compensate for the increased apparent source 
size in the sky image. The mode 06 data added 16\%, 20\%, and 35\%
exposure time for epoch~1b, 3, and 4, respectively. We carefully
checked that there were no significant differences between the mode 01
and 06 data and then combined them using \texttt{addascaspec}.

Although the \nustar- and \suzaku-data of the type II giant
outburst of \gro in 2012 November have been published by
\citet{bellm2014a} already, we analyze these data again for two
reasons. First, our previous analysis of the spectral dependence on
the source's luminosity does not include these observations since
the flux was higher than what had been observed until then.
Secondly, by adding the mode 06 data, our statistical quality is
improved compared to the spectra used by \citet{bellm2014a}.

\nustar-FPMA  and B data were fitted simultaneously between 4--78\,keV.
A first investigation of the spectra showed a significant mismatch
between \swift and  \nustar data below 4\,keV across all epochs, which
is why we did not use \nustar-data down to 3\,keV. This mismatch has
been reported as well by \citet{bellm2014a} for epoch~1b. The
\nustar data were rebinned to a signal-to-noise-ratio of 18 between
4--45\,keV and 6 above 45\,keV. 

The barycentered \nustar-FPMA light curve of \gro during epoch~3 was
extracted from the source region as defined above and with a 1\,s
time resolution.

\subsection{\swift}
\label{sec:data:swift}

The data of the X-Ray Telescope \citep[XRT;][]{burrows2005a,godet2007a}
onboard \swift \citep{gehrels2004a,gehrels2005a} were reprocessed
following the Data Reduction Guide \citep{capalbi2005a}.

The observations of epoch~1b and 4 were operated in the ``Windowed
Timing'' mode (WT) and that of epoch~3 in ``Photon Counting'' mode
(PC). The events of these observations were first calibrated and
screened using \texttt{xrtpipeline} v0.13.2. The applied Telescope
Definition File (teldef) was based on XRT CALDB release 2013-06-01.
After having extracted an image of the observation using
\texttt{xselect}, we created a circular source region with a radius
of $47''$ corresponding to 90\% of the PSF. In WT mode the circular
background region of the same radius was selected at the edge of the
one-dimensional image. During epoch~3 the count rate was,
unfortunately, higher than $0.5\,\mathrm{counts}\,\mathrm{s}^{-1}$,
the nominal limit for pile-up since the outburst did not decay into
quiescence as expected (see Fig.~\ref{fig:batlc}). To avoid pile-up
in PC mode we compared the count rate profile with the expected PSF.
We checked our results with the \texttt{pileest} tool, since the bad
CCD columns due to a micro-meteorite hit might affect the count rate
profile. The final source extraction region is an annulus with an
outer radius of $47''$ and an inner radius of $22''$. The background
region for epoch~3 is an annulus centered at the source position
with an outer radius of $550''$ and an inner radius of $250''$.

The XRT-spectra were extracted using \texttt{xselect}. We only
considered single events (grade 0) in the observations performed in PC
mode (epoch~3). For the remaining observations in WT mode we also
considered split events (grades 0--2). The spectra for each Good Time
Interval (GTI) in WT mode were added into a final single spectrum. The
ARF was generated by \texttt{xrtmkarf} taking the effective area and
filter transmission into account and correcting for vignetting and
PSF.

We rebinned the \swift-XRT spectra to a SNR of 12 for epochs 1b and
4, and to 6 for epoch 3, in order to retain enough spectral
resolution around the iron line. These spectra were fitted in the
1--8\,keV range.

\section{Spectral Analysis}
\label{sec:spec}

The spectral analysis was performed using the Interactive Spectral
Interpretation System \citep[ISIS,][]{houck2000a} v1.6.2-30. Unless
stated otherwise, all uncertainties are given at the 90\% confidence
level and represent single parameter uncertainties ($\Delta \chi^2 =
2.71$).

As found in our previous work \citepalias{kuehnel2013a}, the \rxte
broad-band spectra of \gro can be well described by a power-law with
an exponential cutoff and an additional black body. This model
provides very good fits over more than two orders of magnitudes in the
source's flux. We applied the same model to the recent observations
summarized in Table~\ref{tab:obs} in order to check and extend the
covered range of fluxes. In \citetalias{kuehnel2013a} we have found
the temperature of the black body $kT=1.833 \pm
0.019$\,keV\footnote{The uncertainties of some of the flux independent
  parameters listed in Table~4 of \citetalias{kuehnel2013a} have been
  accidentally given on a confidence level slightly smaller than the
  stated 90\%. The affected parameters with correct uncertainties are
  the black body temperature $kT=1.833 \pm 0.019$\,keV, the folding
  energy $E_\mathrm{fold}=15.9 \pm 0.3$\,keV, the flux of the galactic
  ridge X-ray emission $F_\text{3--10\,keV}=4.25 \pm 0.23 \times
  10^{-12}$\,erg\,s$^{-1}$\,cm$^{-2}$, and its iron line parameters
  $E=6.35 \pm 0.03$\,keV, $\sigma=0.53 \pm 0.06$\,keV, and $F=2.39 \pm
  0.18 \times 10^{-4}$\,photons\,s$^{-1}$\,cm$^{-2}$.} and the folding
energy of the cutoff power-law $E_\mathrm{fold} = 15.9 \pm
0.3$\,keV\footnotemark[1] to be independent of the source's flux and
consistent among the outbursts. Furthermore, the bolometric flux of
the black body, $F_\mathrm{BB}$, and the photon index of the
power-law, $\Gamma$, were found to be well-defined functions of the
source's 15--50\,keV flux, $F_\mathrm{PL}$. When checking these
earlier results against the recent data analyzed here, we did not
apply any of these findings a priori to the following spectral
analysis.

We modeled the absorption by the interstellar medium (ISM) using
\texttt{TBnew}\footnote{\url{http://pulsar.sternwarte.uni-erlangen.de/wilms/research/tbabs/}},
a revised version of the model described by \citet{wilms2000a}.
Element abundances were set to \citet{wilms2000a} and cross-sections
were taken from \citet{verner1996a}. Furthermore, \gro shows
fluorescent emission of iron at 6.4\,keV and a CRSF around 78\,keV,
which we described by a Gaussian and a pseudo-Lorentzian absorption
profile \citep[\texttt{CYCLABS},][]{makishima1990a},
respectively. Finally, we accounted for uncertainties in the flux
calibration using a multiplicative constant, $c_\mathrm{x}$, for each
instrument.

\subsection{\suzaku (epoch 1a and 2)}
\label{sec:spec:suzaku}

For the analysis of the \suzaku-spectra, all broad-band continuum
parameters were allowed to vary, which are the black body temperature,
$kT$, the folding energy, $E_\mathrm{fold}$, the power-law photon
index, $\Gamma$, the black body flux, $F_\mathrm{BB}$, and the overall
flux, $F_\mathrm{PL}$, in the 15--50\,keV range. The absorption column
density, $N_\mathrm{H}$, the energy of the iron K$\alpha$ line,
$E_\mathrm{Fe\,K\alpha}$, and its flux, $F_\mathrm{Fe\,K\alpha}$, were
free parameters as well. The width of the iron line was kept narrow,
i.e, fixed to $\sigma_\mathrm{Fe\,K\alpha} = 10^{-6}$\,keV. From the
best-fit we derived the iron line equivalent width,
$\mathrm{EW}_\mathrm{Fe\,K\alpha}$. The flux calibration constants,
$c_\mathrm{XIS0}$, $c_\mathrm{XIS1}$, $c_\mathrm{PIN}$, and
$c_\mathrm{GSO}$ were determined relative to $c_\mathrm{XIS3}=1$
(fixed). Note that during epoch~1a and 2 further calibration
features were included in the fit as well (see
Sect.~\ref{sec:data:suzaku}). The CRSF was included in modeling
epoch 1 only as it is not detected in epoch 2 (see the following for
details).

\begin{figure}
  \includegraphics[width=\columnwidth]{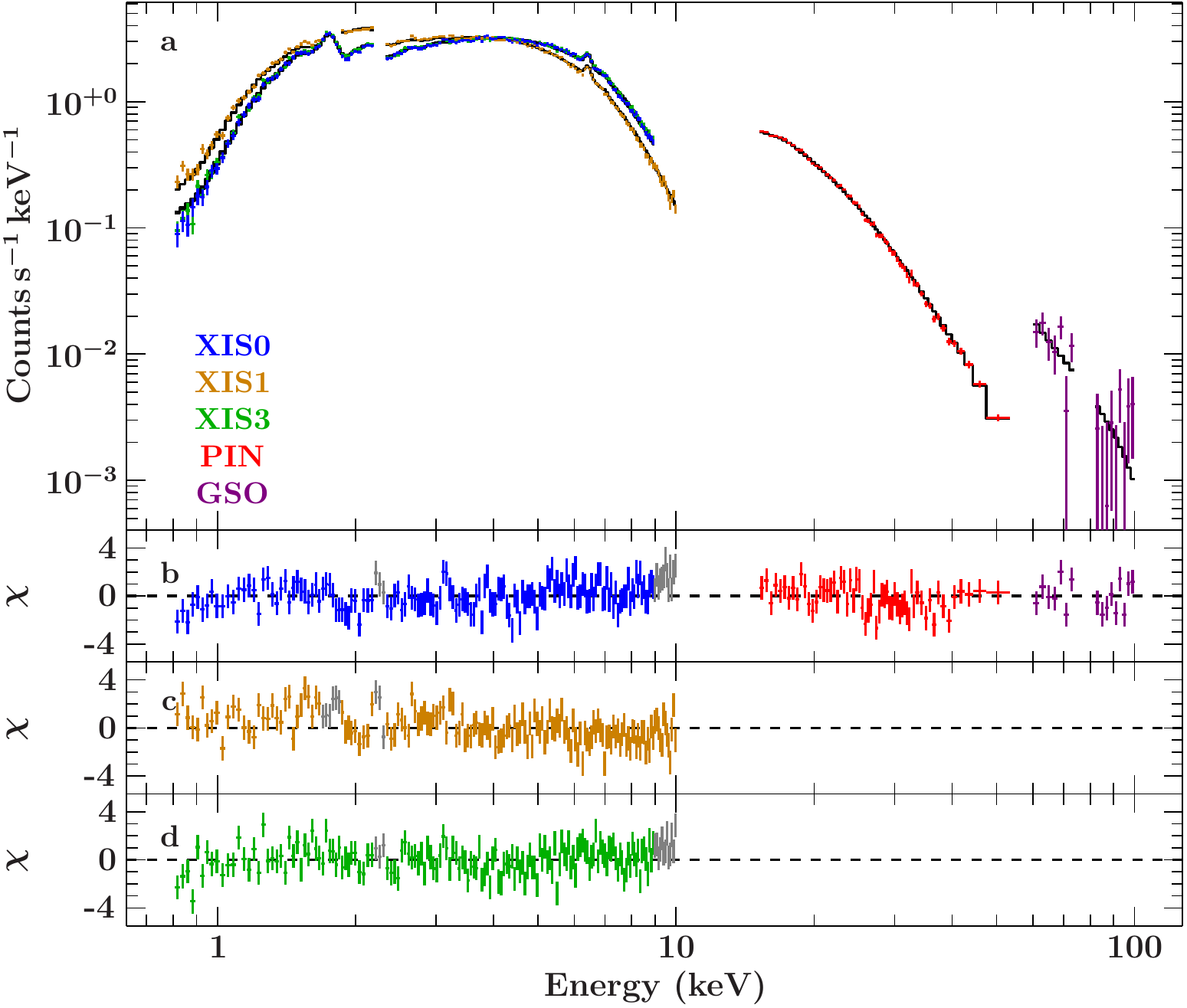}
  \caption{\textbf{a)} The \suzaku-XIS (0: blue; 1: yellow; 3: green),
    -PIN (red), and -GSO (purple) spectra of the 2014 January
    observation of \gro (epoch~2) together with the best-fit model
    (black). \textbf{b-d)} The residuals to the best-fit model (b:
    XIS0, PIN, and GSO; c: XIS1; d: XIS3) showing the ignored energy
    bins due to calibration uncertainties (gray). The residuals of
    XIS1 (c) are shown without taking the photon index shift, $\Delta
    \Gamma$, into account.}
  \label{fig:suz201401}
\end{figure}

A first fit to the data of the 2014 January outburst (epoch~2)
results in a very good description of the data. We found, however, a
line like feature around 7\,keV, which could be modeled by adding a
second narrow Gaussian component. The fitted line energy of
$E_\mathrm{FeK\beta} = 6.85^{+0.21}_{-0.09}$\,keV and relative flux
$F_\mathrm{FeK\beta} / F_\mathrm{FeK\alpha} = 0.19^{+0.10}_{-0.10}$ are in excellent agreement
with fluorescent K$\beta$ emission from neutral iron. The
\suzaku-spectrum together with the final best-fit model is shown in
Fig.~\ref{fig:suz201401}. The corresponding fit parameters are listed
in Table~\ref{tab:bestfits}.

The \suzaku-spectrum during the giant 2012 November outburst
(epoch~1a) showed even more features. After applying the same
\citetalias{kuehnel2013a} model as for epoch~2, strong residuals in
absorption in PIN and GSO at energies above 60\,keV remained, which we
attributed to the high-energy CRSF
\citep{bellm2014a,yamamoto2013a,yamamoto2014a}. Adding such a feature
with its width, $W_\mathrm{cyc}$, fixed to 10\,keV improved the fit
significantly ($\Delta \chi^2 = 353$). The energy and depth of the
CRSF, $E_\mathrm{cyc}$ and $\tau_\mathrm{cyc}$, respectively, were
consistent with those found earlier. The XIS-spectra showed an excess
towards lower energies, which could be described with a second black
body at $kT_2 = 0.489^{+0.028}_{-0.033}$\,keV and improved the fit further
($\Delta \chi^2 = 456$). Finally, the emission lines of iron at 6.40,
6.67, and 7.00\,keV, which we have detected in our earlier analysis of
this dataset \citepalias{kuehnel2013a}, were again detected. The line
at 6.67\,keV represents K$\alpha$ emission from He-like iron
(\ion{Fe}{XXIV}), while the line at 7.00\,keV is a combination of the
K$\beta$ from neutral iron and K$\alpha$ from H-like iron
(\ion{Fe}{XXV}). The total flux in the fitted line was significantly
higher than what is expected from neutral K$\beta$ alone. The fit
parameters of the final best-fit are listed in
Table~\ref{tab:bestfits}. For a Figure showing the full
\suzaku-spectrum see \citet[][Fig.~4]{bellm2014a} and
\citetalias[][Fig.~15]{kuehnel2013a} for a close-up of the iron line
region.

\subsection{\nustar and \swift (epoch~1b, 3, and 4)}
\label{sec:spec:nustar}

In a first step, we applied again the \citetalias{kuehnel2013a} model
to the \nustar- and \swift-spectra (see Sect.~\ref{sec:spec:suzaku}).
The broad-band continuum parameters were kept free as for the \suzaku
analysis above. We added a Gaussian component in order to model the
apparent neutral iron line with its centroid energy,
$E_\mathrm{Fe\,K\alpha}$, its flux and equivalent width, 
$F_\mathrm{Fe\,K\alpha}$ and $EW_\mathrm{Fe\,K\alpha}$, respectively,
and its width, which we fixed to $\sigma_\mathrm{Fe\,K\alpha} =
10^{-6}$\,keV. Flux calibration constants, $c_\mathrm{FPMB}$ and
$c_\mathrm{XRT}$, relative to $c_\mathrm{FPMA}=1$ were included during
the fit.

\begin{figure*}
\begin{minipage}[t]{.485\textwidth}
  \includegraphics[width=\columnwidth]{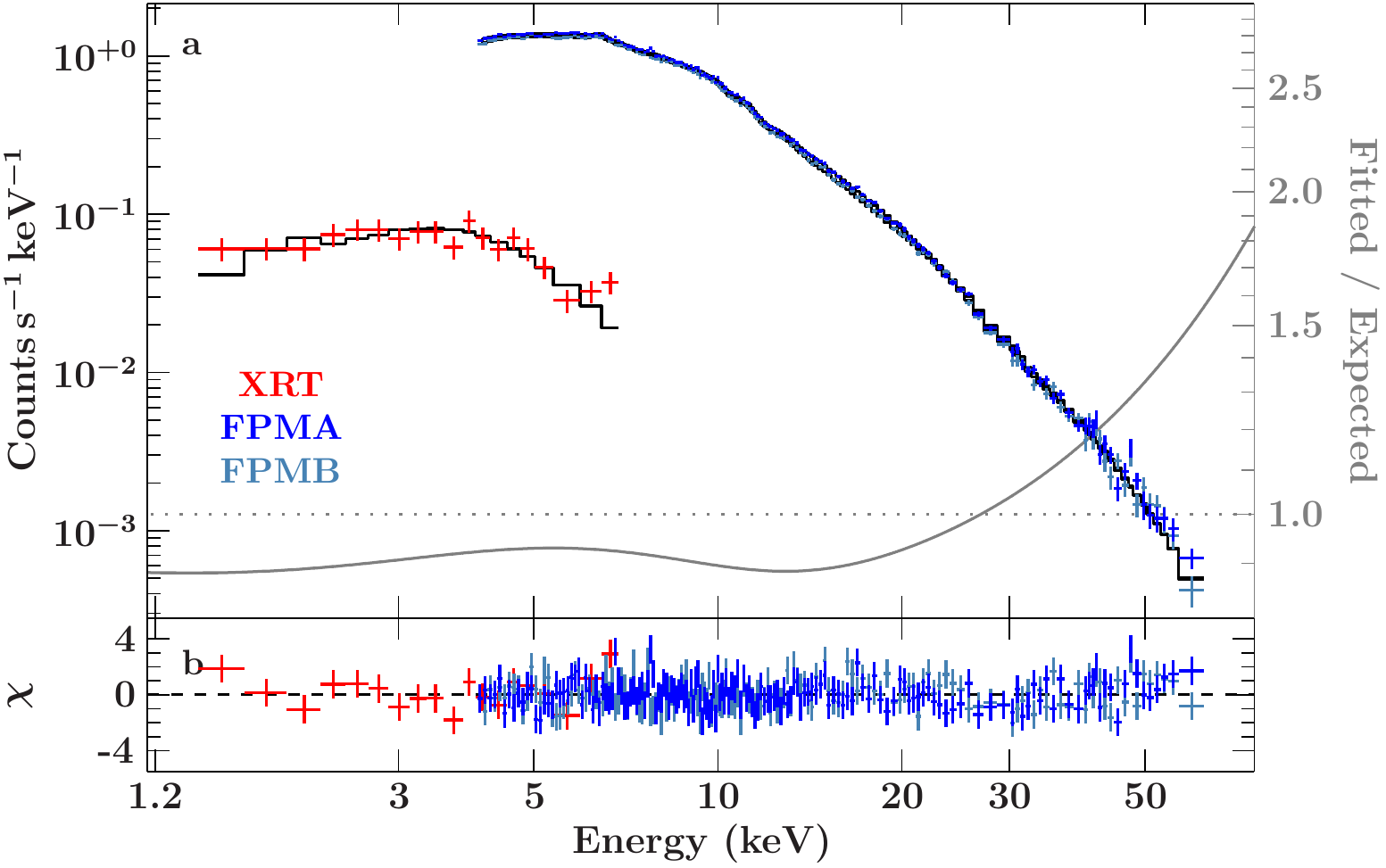}
  \caption{\textbf{a)} \nustar-FPMA (dark blue), -FPMB (light blue),
    and \swift-XRT spectrum (red) of \gro in 2014 December after the
    second peak of its triple peaked outburst (epoch~3). The best-fit
    model is shown in black and the gray line shows its ratio to the
    spectral shape as predicted by our spectral evolution model (see
    Sect.~\ref{sec:discuss:parevol} and
    Fig.~\ref{fig:parfluxcorr}). The data have been rebinned for
    display purposes. \textbf{b)} The residuals of the best-fit
    model.}
  \label{fig:nus201412}
\end{minipage}\hfill\begin{minipage}[t]{.485\textwidth}
  \includegraphics[width=\columnwidth]{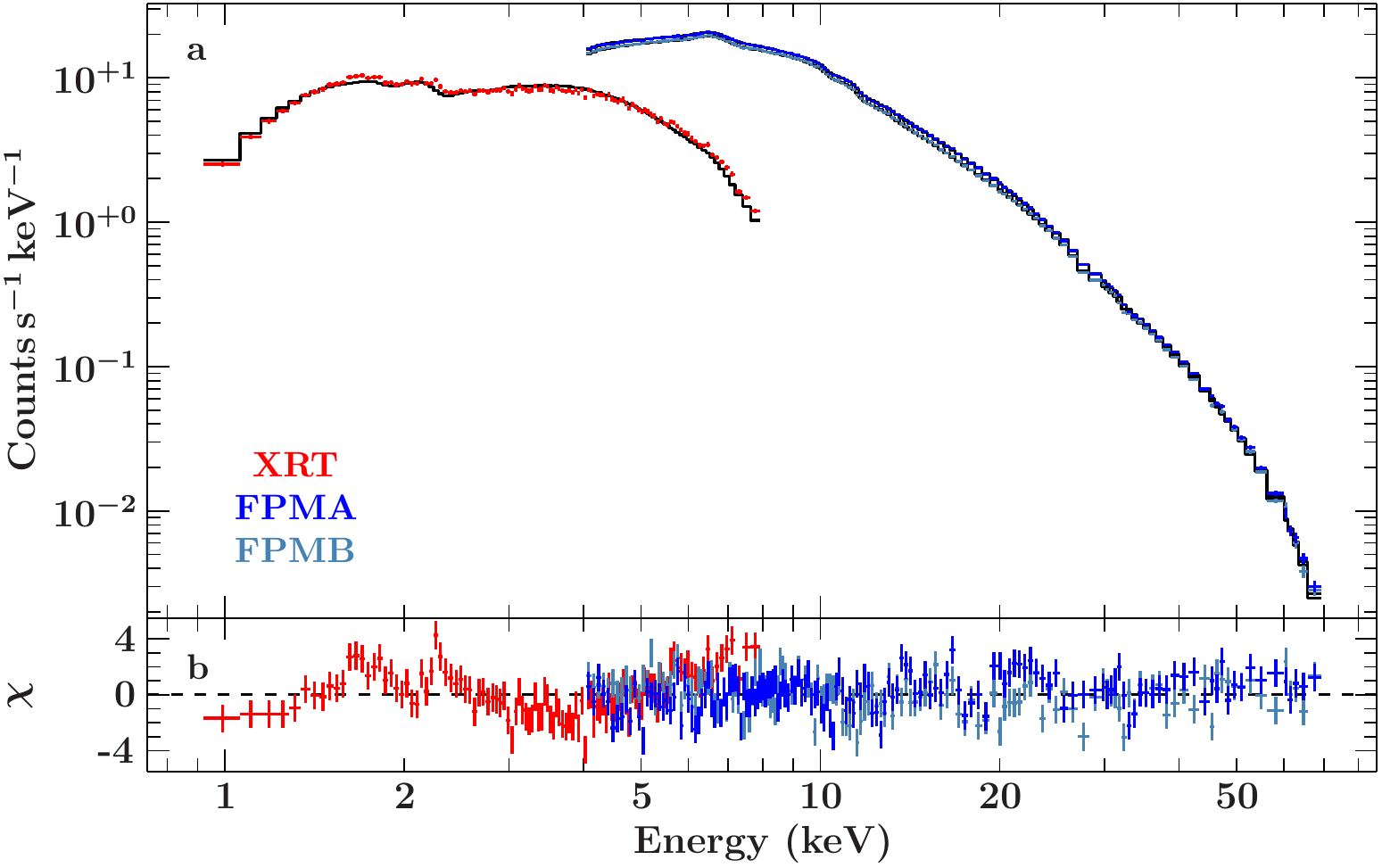}
  \caption{\textbf{a)} \nustar-FPMA (dark blue), -FPMB (light
    blue), and \swift-XRT spectrum (red) of \gro in 2015 January
    shortly after maximum flux of the third peak of the triple
    peaked outburst (epoch~4). The black line shows the best-fit
    model. The data have been rebinned for display purposes.
    \textbf{b)} The residuals of the best-fit model.}
  \label{fig:nus201501}
\end{minipage}
\end{figure*}

This model was able to describe the spectra taken during epoch~3,
when \gro's luminosity was just above the detection limit of
\swift-BAT (compare Fig.~\ref{fig:batlc}). The spectra during high
luminosities of the source (epochs~1b and 4) could not, however, be
fitted well with this model. There were clear indications for the CRSF
towards the higher end of \nustar's energy range. Including a CRSF
with its width fixed to $W_\mathrm{cyc} = 10$\,keV as for the
\suzaku-data (see Sect.~\ref{sec:spec:suzaku}) led to a very good
description of the spectra ($\Delta \chi^2 = 527$ for epoch~1b and
$\Delta \chi^2 = 286$ for epoch~4). Slight residuals around the
neutral iron line remained in all epochs. The energy resolution of the
\nustar-FPMs, however, does not allow resolution of further emission
lines, such as iron K$\beta$ or ionized iron, as was possible with the
\suzaku-XISs. Nevertheless, varying the width of the iron K$\alpha$
line, $\sigma_\mathrm{Fe\,K\alpha}$, was sufficient to get rid of
these residuals. The best-fit energy, $E_\mathrm{Fe\,K\alpha}$, was
higher than the 6.4\,keV expected from neutral iron. Together with a
width of a few 100\,eV these results point towards a blend of several
emission lines, possibly from ionized iron as has been detected in the
\suzaku-data (see Sect.~\ref{sec:spec:suzaku}). For epoch~1b, the flux
in the line of around 75\,ph\,s$^{-1}$\,cm$^{-2}$ agrees with the sum
of the fluxes in the individual lines resolved in the \suzaku-XIS
spectra (epoch~1a). The final fit-parameters for all epochs of
\nustar- and \swift-observations are listed in
Table~\ref{tab:bestfits}. The spectra and corresponding models for
epoch~3 and 4 are shown in Fig.~\ref{fig:nus201412} and
\ref{fig:nus201501}, respectively. For epoch~1b see
\citet[][Fig.~3]{bellm2014a}. Note that the line-like residuals in
Fig.~\ref{fig:nus201501}b at ${\sim}1.8$\,keV and ${\sim}2.2$\,keV are
only visible due to the coarser channel binning used in this figure.
These features are calibration uncertainties in \swift-XRT around the
Si K edge and the Au M edge, respectively
\citep{godet2007a,hurkett2008a}. Due to their low significance they do
not have any effect on the spectral parameters here.

\begin{table*}
  \caption{Best-fit parameters of the \suzaku- and simultaneous
  \nustar-\swift-spectra analyzed here. The uncertainties are given
  at the 90\% confidence level. The parameters regarding the
  calibration of the XISs and PIN are given in
  Sect.~\ref{sec:spec:suzaku}.}
  \label{tab:bestfits}
  \centering
  \begin{tabular}{llllll}
  \hline\hline
          & Epoch 1a & Epoch 1b & Epoch 2 & Epoch 3 & Epoch 4 \\
Parameter & \suzaku  & \nustar-\swift & \suzaku & \nustar-\swift & \nustar-\swift \\

  \hline\\[-1.9ex]
  ${\chi^2}_\mathrm{red} / \mathrm{d.o.f.}$ & 1.224 / 579 & 1.180 / 1808 & 1.138 / 525 & 0.956 / 579 & 1.175 / 1796 \\[2pt]
$F_\mathrm{PL}$ ($10^{-9}\,$erg\,s$^{-1}$cm$^{-2}$ [15--50\,keV]) & $10.9^{+0.9}_{-0.9}$ & $10.357^{+0.024}_{-0.024}$ & $1.49^{+0.08}_{-0.08}$ & $0.3070^{+0.0030}_{-0.0030}$ & $7.121^{+0.018}_{-0.018}$ \\[2pt]
$\Gamma$ & $0.32^{+0.16}_{-0.25}$ & $0.476^{+0.024}_{-0.025}$ & $0.63^{+0.05}_{-0.05}$ & $1.15^{+0.08}_{-0.08}$ & $0.582^{+0.020}_{-0.021}$ \\[2pt]
$E_\mathrm{fold}$ (keV) & $14.3^{+1.8}_{-1.9}$ & $14.86^{+0.25}_{-0.24}$ & $15.9^{+0.8}_{-0.8}$ & $19.6^{+1.5}_{-1.4}$ & $15.75^{+0.26}_{-0.25}$ \\[2pt]
$F_\mathrm{BB}$ ($10^{-9}\,$erg\,s$^{-1}$cm$^{-2}$) & $1.2^{+0.5}_{-0.4}$ & $0.90^{+0.08}_{-0.08}$ & $0.54^{+0.04}_{-0.04}$ & $0.063^{+0.011}_{-0.010}$ & $0.82^{+0.05}_{-0.05}$ \\[2pt]
$kT$ (keV) & $1.89^{+0.22}_{-0.22}$ & $1.689^{+0.023}_{-0.021}$ & $1.90^{+0.04}_{-0.05}$ & $1.69^{+0.08}_{-0.07}$ & $1.949^{+0.029}_{-0.027}$ \\[2pt]
$kT_2$ (keV) & $0.489^{+0.028}_{-0.033}$ & - & - & - & - \\[2pt]
${F_\mathrm{BB2}/F_\mathrm{BB}}^{\ddag}$ & $0.46^{+0.12}_{-0.09}$ & - & - & - & - \\[2pt]
$N_\mathrm{H}$ ($10^{22}\,$cm$^{-2}$) & $1.45^{+0.07}_{-0.07}$ & $1.44^{+0.05}_{-0.05}$ & $1.60^{+0.04}_{-0.04}$ & $4.0^{+0.7}_{-0.7}$ & $1.66^{+0.05}_{-0.05}$ \\[2pt]
$E_\mathrm{Fe\,K\alpha}$ (keV) & $6.405^{+0.030}_{-0.010}$ & $6.528^{+0.014}_{-0.014}$ & $6.433^{+0.010}_{-0.021}$ & $6.43^{+0.06}_{-0.04}$ & $6.559^{+0.014}_{-0.014}$ \\[2pt]
$\sigma_\mathrm{Fe\,K\alpha}$ (keV) & $10^{-6}$\,$^\dagger$ & $0.306^{+0.020}_{-0.019}$ & $10^{-6}$\,$^\dagger$ & $10^{-6}$\,$^\dagger$ & $0.292^{+0.019}_{-0.018}$ \\[2pt]
$F_\mathrm{Fe\,K\alpha}$ ($10^{-4}$\,ph\,s$^{-1}$\,cm$^{-2}$) & $26.7^{+3.0}_{-3.0}$ & $73^{+4}_{-4}$ & $9.1^{+0.8}_{-0.8}$ & $1.24^{+0.27}_{-0.27}$ & $53.1^{+2.6}_{-2.5}$ \\[2pt]
$EW_\mathrm{Fe\,K\alpha}$ (eV) & $36^{+5}_{-5}$ & $102^{+5}_{-5}$ & $50^{+5}_{-5}$ & $26^{+4}_{-4}$ & $90^{+5}_{-5}$ \\[2pt]
$E_\mathrm{Fe\,K\beta}$ (keV) & ${7.000^{+0.023}_{-0.058}}\,^{\star}$ & - & $6.85^{+0.21}_{-0.09}$ & - & - \\[2pt]
$F_\mathrm{Fe\,K\beta}$ ($10^{-4}$\,ph\,s$^{-1}$\,cm$^{-2}$) & ${11.1^{+3.0}_{-3.0}}\,^{\star}$ & - & $1.6^{+0.8}_{-0.8}$ & - & - \\[2pt]
$EW_\mathrm{Fe\,K\beta}$ (eV) & ${16^{+5}_{-5}}\,^{\star}$ & - & $10^{+5}_{-5}$ & - & - \\[2pt]
$E_\mathrm{\ion{Fe}{XXIV}}$ (keV) & $6.689^{+0.018}_{-0.022}$ & - & - & - & - \\[2pt]
$F_\mathrm{\ion{Fe}{XXIV}}$ ($10^{-4}$\,ph\,s$^{-1}$\,cm$^{-2}$) & $23.6^{+3.0}_{-3.0}$ & - & - & - & - \\[2pt]
$EW_\mathrm{\ion{Fe}{XXIV}}$ (eV) & $33^{+5}_{-5}$ & - & - & - & - \\[2pt]
$E_\mathrm{cyc}$ (keV) & $75.9^{+1.6}_{-1.4}$ & $70.7^{+1.7}_{-1.5}$ & - & - & $70.4^{+2.7}_{-1.9}$ \\[2pt]
$W_\mathrm{cyc}$ (keV) & 10$^\dagger$ & $10^\dagger$ & - & - & $10^\dagger$ \\[2pt]
$\tau_\mathrm{cyc}$ & $1.8^{+0.5}_{-0.5}$ & $0.99^{+0.18}_{-0.13}$ & - & - & $0.75^{+0.24}_{-0.14}$ \\[2pt]
$c_\mathrm{XIS0} / c_\mathrm{XIS3}$ & $0.980^{+0.005}_{-0.005}$ & - & $0.943^{+0.006}_{-0.006}$ & - & - \\[2pt]
$c_\mathrm{XIS1} / c_\mathrm{XIS3}$ & $1.046^{+0.005}_{-0.005}$ & - & $0.869^{+0.012}_{-0.012}$ & - & - \\[2pt]
$c_\mathrm{PIN} / c_\mathrm{XIS3}$ & $1.38^{+0.12}_{-0.11}$ & - & $1.31^{+0.07}_{-0.07}$ & - & - \\[2pt]
$c_\mathrm{GSO} / c_\mathrm{XIS3}$ & $1.43^{+0.15}_{-0.13}$ & - & $1.20^{+0.23}_{-0.22}$ & - & - \\[2pt]
$c_\mathrm{FPMB} / c_\mathrm{FPMA}$ & - & $1.0269^{+0.0015}_{-0.0015}$ & - & $1.034^{+0.005}_{-0.005}$ & $1.0147^{+0.0015}_{-0.0015}$ \\[2pt]
$c_\mathrm{XRT} / c_\mathrm{FPMA}$ & - & $1.187^{+0.009}_{-0.009}$ & - & $1.00^{+0.08}_{-0.08}$ & $1.233^{+0.010}_{-0.010}$ \\[2pt]

  \hline
  \end{tabular}
  \tablefoot{
  \tablefoottext{$\ddag$}{Ratio between the bolometric fluxes of the black bodies.}
  \tablefoottext{$\star$}{Epoch 1a is a blend of FeK$\beta$ and
  \ion{FeK$\alpha$}{XXV}.}
  \tablefoottext{$\dagger$}{Fixed.}
  }
\end{table*}

\section{Discussion on Individual Observations}
\label{sec:discuss}

We have successfully applied the model which was used in our
previous study of \gro \citepalias{kuehnel2013a}, to recent \suzaku-
and simultaneous \nustar- and \swift-spectra. We compare the results of
\citetalias{kuehnel2013a}, especially the parameter evolution with the
X-ray flux, with the results of the spectral analysis presented here in
Sect.~\ref{sec:discuss:parevol} separately. In this section we discuss
our results in their individual context.

\subsection{CRSF and 2$^{\rm nd}$ Black Body}

During the giant 2012 November outburst we can confirm the presence of
a CRSF in the \suzaku- and \nustar-data (epochs~1a and
1b). Furthermore, we have detected the CRSF in the \nustar-data during
epoch~4, which happened at a comparable source luminosity level as
during epoch~1a and 1b. There is no hint of a cyclotron line in the
\nustar-data during epoch~3 and in the \suzaku-data during epoch~2,
when the source was at a much lower luminosity level. We cannot,
however, exclude its presence due to the lower statistic. In the
\suzaku-data of the giant outburst (epoch~1a), we find the cyclotron
line energy $E_\mathrm{cyc}=\ $$75.9^{+1.6}_{-1.4}$\,keV to be in
excellent agreement with $75.5^{+2.5}_{-1.5}$\,keV and
$78^{+3}_{-2}$\,keV as found in the earlier analyses of these data by
\citet{yamamoto2013a,yamamoto2014a} and \citet{bellm2014a},
respectively. The apparent difference in the measured CRSF energies
between \nustar (epoch 1b and 4) and \suzaku (epoch 1a) is likely due
to systematic uncertainties in this energy range, which is at the
upper end of the useful energy range for both
instruments. Furthermore, a slight change in the continuum parameters,
especially the photon index, $\Gamma$, or the folding energy,
$E_\mathrm{fold}$, influences the CRSF parameters at this high
energy. Thus, we cannot draw definite conclusions from the apparent
difference between the \suzaku- (epoch~1a) and the \nustar-results
(epochs~1b and 4).

In the \suzaku-data of the giant outburst of \gro in 2012
(epoch~1a) a second black body with $kT_2 = 0.489^{+0.028}_{-0.033}$\,keV is
required to fit the low energy part of the XIS-spectra successfully.
We had already detected this feature in our earlier analysis of these
data together with \swift-spectra \citepalias{kuehnel2013a}. On the
other hand, the feature was not detected in the two \suzaku
observations at lower luminosities (epoch~2 and during the decay of
the 2007 data, see \citetalias{kuehnel2013a} for details) as well as
in the \swift data during the maximum of the 2007 outburst
\citepalias[see][]{kuehnel2013a}. Thus, it is likely that the soft
part of \gro's spectrum at $\lesssim$3\,keV changed during the
giant outburst. In combination with the evolution of the other
spectral parameters with flux, as presented
Sect~\ref{sec:discuss:parevol}, this is evidence for the existence of
different accretion regimes in \gro. \citet{bellm2014a}, who used the
NPEX model for the broad-band continuum, found this second black body
at the same temperature as we did, while they did not detect it in the
\nustar data alone. We found that this feature is not significantly
detected in the combined \nustar- and \swift-data as well. The
broad-band continuum parameters agree, however, between \suzaku and
\nustar. Thus, the presence of this feature does not influence the
remaining spectral parameter much. Adding this feature to the \nustar
spectrum at a fixed temperature indeed leads to consistent results
with the \suzaku spectrum.

\subsection{Spectral Anomaly of Epoch 3}
\label{sec:discuss:epochIV}

\begin{figure}
  \includegraphics[width=\columnwidth]{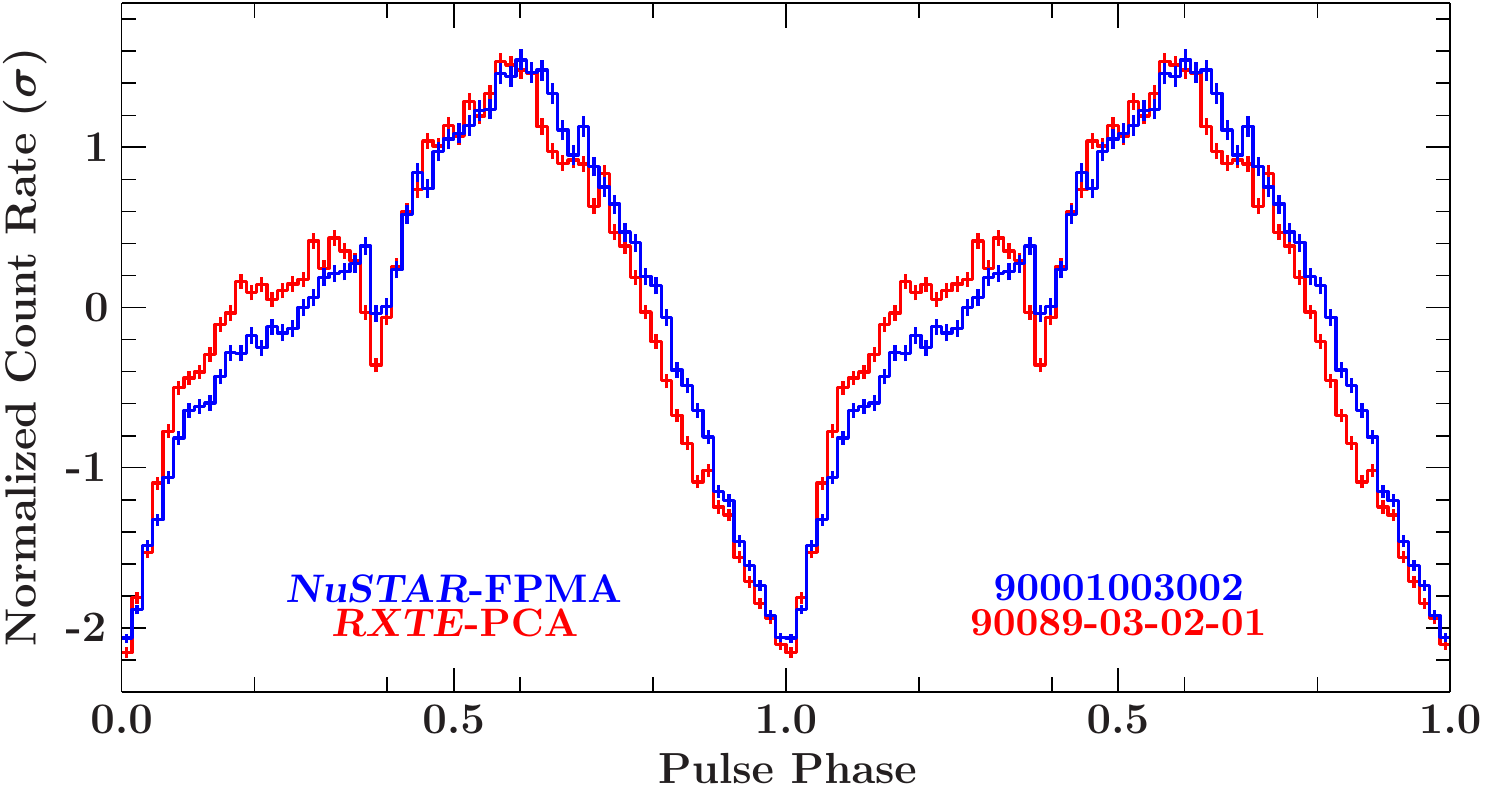}
  \caption{Pulse profiles of \gro during epoch~3 (blue;
  \mbox{\nustar}-FPMA) and 2005 (red; \rxte-PCA). The \rxte observation was
  performed at a similar flux level of the source compared to the
  \nustar observation. The pulse profiles have been normalized to a
  mean count rate of zero and a standard deviation, $\sigma$, of
  unity. Pulse phase zero is defined as the minimum count rate.}
  \label{fig:profilenustar}
\end{figure}

During the transition from the second to the third outburst (epoch~3)
during \gro's ``triple-peaked'' outburst, the spectral parameters
disagree with the correlation between spectral parameters as found
in the other observations of this source (see
Sect.~\ref{sec:discuss:parevol} for details). In particular, the
folding energy, $E_\mathrm{fold}$, of ${\sim}19.6$\,keV is
significantly higher than the 15--16\,keV as found during all other
epochs (see Table~\ref{tab:bestfits}). The gray line in
Fig.~\ref{fig:nus201412}a shows the ratio between the best-fit model
spectrum and the predicted spectrum assuming the measured source flux
at 15--50\,keV as listed in Table~\ref{tab:bestfits}. As can be seen,
this flux ratio increases exponentially above ${\sim}$30\,keV
up to a factor of 2 around 70\,keV. These facts still hold true if the
\nustar and \swift spectra are fitted without taking the CRSF into
account, i.e., the presence or absence of the CRSF has no influence on
the model parameters.

To draw conclusions about the reason for this spectral anomaly we
compared the pulse profile of the corresponding \nustar observation
with an \rxte observation in 2005 at a similar flux level (ObsID
90089-03-02-01). As \rxte-PCA and \nustar-FPMA are sensitive over
comparable energy ranges (3--60 and 3--78\,keV), differences in the
pulse profiles can be associated with changes in the accretion
geometry. After correcting the light curves for binary motion using
the orbital parameter listed in \citetalias{kuehnel2013a}, we folded
the \nustar and \rxte light curves on 93.445\,s and
93.675\,s, respectively, which correspond to the spin periods of
\gro and were determined from the data using the epoch folding
technique \citep[see, e.g.,][]{leahy1983a}. 

The resulting pulse profiles shown in Fig.~\ref{fig:profilenustar}
were normalized to a mean count rate of zero and a standard deviation
of unity \citep[see, e.g.,][]{ferrigno2011a,schoenherr2014a}. Both
pulse profiles are very similar in shape. We therefore
concluded that the accretion geometry between these observations was
very similar as well, which hints at an origin outside of the neutron
star's magnetosphere for the spectral anomaly. The observed changes in
the pulse profile of A0535+26 during pre-outburst flares
\citep{caballero2008a}, were explained by \citet{postnov2008a} as the
result of magnetospheric instabilities. Since the pulse profile of
\gro during the flaring epoch~3 is consistent with the expected pulse
profile shape, it is unlikely that these instabilities are the origin
for the observed flaring activity.

\subsection{``Triple-peaked'' Outburst Morphology}

After the first confirmed giant outburst of \gro in 2012 November
(epochs~1a and~1b) since its discovery in 1993, the source featured an
expected type I outburst in 2014 January (epoch~2). Surprisingly,
after the type I outburst in 2014 September, \gro went into a type II
outburst again.  But instead of fading into quiescence after this
second outburst (epoch~3), the source showed a third outburst in a
single orbit (epoch~4). In fact, the peak of the third outburst was
reached at apastron \citep[see Fig.~\ref{fig:batlc}
and][]{kretschmar2015a}.

Figure~\ref{fig:batlc} shows the light curve morphology of \gro
during this ``triple-peaked'' outburst in 2014/2015, which lasted for
half the orbital period of ${\sim}250$\,d. The transition between
the first two outbursts occurred rather smoothly, which has also
been observed during the known ``double-peaked'' outbursts of
A~0535+262 \citep[][Fig.~1]{caballero2013a} and GX~304$-$1
\citep[][Fig.~1]{postnov2015a}. The peak separation in \gro of
${\sim}65\,\mathrm{d} = 0.25\,P_\mathrm{orb}$ is, however, much
larger than in these systems, which show separations of
${\sim}9\,\mathrm{d} = 0.08\,P_\mathrm{orb}$ with $P_\mathrm{orb} =
111.1$\,d in the case of A~0535+26 \citep{finger2006a} and
${\sim}25\,\mathrm{d} = 0.19\,P_\mathrm{orb}$ with $P_\mathrm{orb} =
132.5$\,d in the case of GX~304$-$1 \citep{priedhorsky1983a}. As
argued by \citet{postnov2015a}, the ``double-peaked'' outburst of
GX~304$-$1 is due to a misaligned Be-disk with respect to the
orbital plane. Once the disk has grown sufficiently beyond the
distance between the stellar surface and the neutron's stars orbit a
second intersection might be possible. This is consistent with
recent theoretical investigations by \citet{okazaki2013a}. The
geometries of the Be-disk and the orbit are unique for each BeXRB,
although there are similarities among this class. Thus, a difference
of the peak separation up to a factor of ${\sim}$3 in orbital phase
is not astonishing. Note that the outbursts of XTE~J1946+274
occurred twice per orbital period \citep[$P_\mathrm{orb}\sim172$\,d;
see, e.g.,][and references therein]{marcu2015b}, i.e., the peak
separation in this system is about 0.5 in orbital phase. As the
system shows $\ge 5$ outbursts in a row, this morphology is, however,
different compared to the ``double-peaked'' outbursts.

Between the second and third peak of \gro's ``triple-peaked''
outburst the light curve is not as smooth as the transition between
the first two peaks. Rather, the source stayed at a more or less
constant level for ${\sim}50$\,d with a weak flaring activity. A
similar behavior was also observed before the giant 2012
November outburst (see Fig.~\ref{fig:batlc}), when the flaring
activity was, however, much stronger and lasted for ${\sim}60$\,d.
These flares occurred quasi-periodically with a period around 9\,d. 
A speculative explanation for this different behavior compared to
the smooth transition might be due to external torques onto the
accretion disk as studied by \citet{dogan2015a}. They argue that
for an accretion disk of a certain size and inclined with respect to
the orbital plane, the torques might overpower the internal disk
torques. Consequently, the disk would break up into slices, which
would precess independently of each other. \citet{dogan2015a} thus
conclude that the mass accretion rate onto the compact object gets
modulated. In their example simulation \citep[see Fig.~5
of][]{dogan2015a} the period of this modulation is much longer than
the orbital period, which makes it difficult to reconcile with the
observed 9\,d period.

\section{Parameter Evolution with Flux}
\label{sec:discuss:parevol}

In \citetalias{kuehnel2013a} we have found that the broad-band
continuum of \gro is a function of the overall X-ray flux. The black
body temperature, $kT$, and the folding energy,
$E_\mathrm{fold}$\,keV, are independent of the 15--50\,keV flux,
$F_\mathrm{PL}$, and consistent among the outbursts. The power-law
photon index, $\Gamma$, and the black body flux, $F_\mathrm{BB}$, show
a tight correlation with $F_\mathrm{PL}$. These results are mainly
based on observations taken by \rxte with the addition of one \swift
and one \suzaku pointing.

The black body temperature found by analyzing the new \suzaku- (epochs~1a
and~2) and joint \nustar- and \swift-observations (epochs 1b, 3, and
4) are all within 0.15\,keV around $kT = 1.833\pm0.019$\,keV as found
by \citetalias{kuehnel2013a}. Despite this small temperature range,
the value during the bright \nustar observations (epochs 1b and 4)
are, however, significantly lower than the mean value. The folding
energy of $E_\mathrm{fold} = 15.92^{+0.29}_{-0.30}$\,keV we have
measured previously in \citetalias{kuehnel2013a} is consistent with
the observations analyzed here with the exception of the
\nustar-epochs 1b and 3. While \gro's spectral anomaly during epoch 3
has been discussed in Sect.~\ref{sec:discuss:epochIV},
$E_\mathrm{fold}$ during epoch 1b is ${\sim}1$\,keV lower compared to
the \rxte result of \citetalias{kuehnel2013a}. In order to understand
these apparent differences in $kT$ and $E_\mathrm{fold}$ we have
investigated contour maps of these parameters against other continuum
parameters. We discovered parameter degeneracies, especially between
the photon index, $\Gamma$, and the folding energy, $E_\mathrm{fold}$.
In fact, the folding energy found previously by
\citetalias{kuehnel2013a} is within 2.7$\sigma$ (for two degrees of
freedom) of the best-fit of epoch~1b as listed in
Table~\ref{tab:bestfits}. Fixing $E_\mathrm{fold} = 15.92$\,keV indeed
leads to a similar goodness of the fits for all epochs (except epoch~3
as discussed above). Thus, we conclude that the flux independent
parameters still seem to be the same for the newer data analyzed here
(except epoch~3), which confirms our previous results
\citepalias{kuehnel2013a}.

\begin{figure}
  \includegraphics[width=\columnwidth]{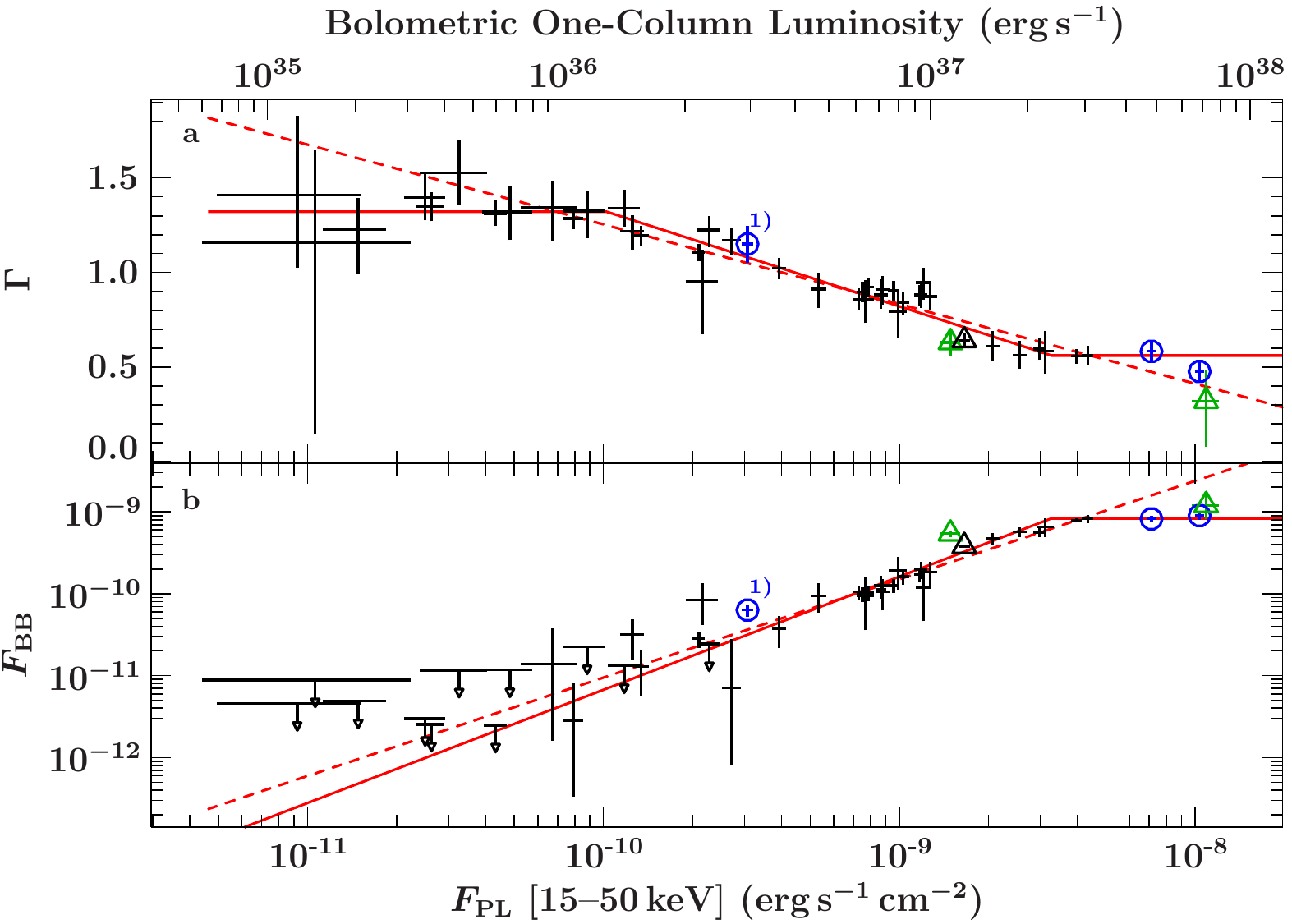}
  \caption{Evolution of the spectral parameters with the 15--50\,keV
    flux, $F_\mathrm{PL}$, of \gro. \textbf{a)} The power-law photon
    index, $\Gamma$, and \textbf{b)} the black body flux
    $F_\mathrm{BB}$ (in
    $\mathrm{erg}\,\mathrm{s}^{-1}\,\mathrm{cm}^{-2}$), with the
    source's flux (uncertainties represent $1\sigma$ uncertainties
    with two degrees of freedom). The black data points are the
    results of our previous \rxte- and the black triangle the joined
    \suzaku-\rxte-analysis \citepalias{kuehnel2013a}. Their fluxes
    have been corrected for calibration uncertainties (see the text
    for details). The new results presented in this paper are the
    green triangles based on \suzaku- and the blue circles based on
    \nustar-data. The red lines show the empirical model with (solid)
    and without (dashed) the breaks at certain luminosity levels as
    described in the text. The \nustar-results marked with 1) have
    been excluded from these fits. The bolometric luminosity for one
    accretion column has been calculated as described in
    Sect.~\ref{sec:regimes}.}
  \label{fig:parfluxcorr}
\end{figure}

In \citetalias{kuehnel2013a} we have also discovered that the photon
index, $\Gamma$, and the black body flux, $F_\mathrm{BB}$, are
functions of the overall 15--50\,keV source flux, $F_\mathrm{PL}$. In
particular, the same behavior is found in all outbursts.
Figure~\ref{fig:parfluxcorr} is an updated version of our previous
\rxte results \citepalias[Fig.~10 of][]{kuehnel2013a}, where the
best-fit $\Gamma$ and $F_\mathrm{BB}$ over $F_\mathrm{PL}$ of the
newer data have been added (see Table~\ref{tab:bestfits}). Note that
since two fit parameters are plotted against each other in each panel,
the uncertainties have to be calculated assuming two degrees of
freedom. As we will fit these parameters below, we are providing
$1\sigma$ uncertainties ($\Delta \chi^2=2.30$).

The parameters of the \suzaku epoch~2 (green triangle at
${\sim}1.5\times10^{-9}\,\mathrm{erg}\,\mathrm{s}^{-1}\,\mathrm{cm}^{-2}$)
are consistent with the behavior seen earlier (black triangle).
Furthermore, the \nustar and \suzaku observations during the giant
2012 outburst (the most luminous blue circle and green triangle) also
give consistent results, although a second black body was necessary to
describe the \suzaku data. Interestingly, the recent \nustar
observation during the third outburst of the ``triple-peaked''
outburst (blue circle at
${\sim}7\times10^{-9}\,\mathrm{erg}\,\mathrm{s}^{-1}\,\mathrm{cm}^{-2}$)
results in the same parameters as for the 2012 giant outburst,
although the source was ${\sim}45\%$ brighter then. The data from
epoch~3 (blue circle marked with 1), where the folding energy,
$E_\mathrm{fold}$, was not consistent with all the other observations
(see Sect.~\ref{sec:discuss:epochIV} for a discussion), is almost
consistent with the apparent parameter evolution. Due to the
inconsistency in $E_\mathrm{fold}$ we, however, ignore this
observation until the end of this section.

In the following, we model the apparent evolution of the power-law
photon index, $\Gamma$, and the black body flux, $F_\mathrm{BB}$,
with the source flux, $F_\mathrm{PL}$, using phenomenological
functions,
\begin{align}\label{eq:corrfuns1}
  {\rm log\!}: X(F_\mathrm{PL}) &= a_X + b_X \log(F_\mathrm{PL} /
  10^{-9}\,\mathrm{erg}\,\mathrm{s}^{-1}\,\mathrm{cm}^{-2}), \\
  {\rm lin\!}: X(F_\mathrm{PL}) &= b_X (F_\mathrm{PL} /
  10^{-9}\,\mathrm{erg}\,\mathrm{s}^{-1}\,\mathrm{cm}^{-2} - a_X), \\
  {\rm pow\!}: X(F_\mathrm{PL}) &= a_X (F_\mathrm{PL} /
  10^{-9}\,\mathrm{erg}\,\mathrm{s}^{-1}\,\mathrm{cm}^{-2})^{b_X},
  \label{eq:corrfuns2}
\end{align}
where $X$ is either the photon index, $\Gamma$, or the black body
flux, $F_\mathrm{BB}$. In case of fitting the evolution of the black
body flux, i.e, $X=F_\mathrm{BB}$ we restrict the black body to be in
emission only, i.e., $F_\mathrm{BB}(F_\mathrm{PL}) \ge 0$ for all
source fluxes, $F_\mathrm{PL}$. In order to fit a model consisting of
these functions to one of the parameter evolutions shown in
Fig.~\ref{fig:parfluxcorr} we minimize the $\chi^2$ accounting for the
asymmetric uncertainties in both the flux, $F_\mathrm{PL}$, and the
parameter of interest, $X$. 

\begin{table}
  \caption{Parameters of the best-fitting combination of models
  describing the spectral evolution as shown in
  Fig.~\protect\ref{fig:parfluxcorr}. The definitions of the models
  are given in the text.}
  \label{tab:corrmodels}
  \centering
  \begin{tabular}{ccc}
  \hline\hline
   & $\Gamma(F_\mathrm{PL})$ & $F_\mathrm{BB}(F_\mathrm{PL})$ \\ & dbl-brkn log & high-brkn pow \\
\hline\\[-1.9ex]
$a_X$ & $0.821^{+0.019}_{-0.019}$ & $0.162^{+0.013}_{-0.013}$\\[2pt]
$b_X$ & $-0.51^{+0.05}_{-0.05}$ & $1.38^{+0.12}_{-0.11}$\\[2pt]
$F_\mathrm{brk,lo}$ & $0.10^{+0.05}_{-0.04}$ & -\\[2pt]
$F_\mathrm{brk,hi}$\tablefootmark{a} & \multicolumn{2}{c}{$3.27^{+0.33}_{-0.30}$}\\[2pt]
${\chi^2}\,^{b}$ & 36.41 & 44.81\\[2pt]
$\chi^2_\mathrm{red}$ / d.o.f.$^{c}$ & \multicolumn{2}{c}{0.97 / 84}\\[2pt]

  \hline
  \end{tabular}
  \tablefoot{
    \tablefoottext{a}{$F_\mathrm{brk,hi}$ has been
    determined by a joint fit to both correlations.}
    \tablefoottext{b}{The $\chi^2$ sum over all 45 data points for
    each parameter evolution.}
    \tablefoottext{c}{The total goodness of the fit.}
  }
\end{table}

During a first investigation of the parameter evolutions, we have ignored
data at
$F_\mathrm{PL}<8\times10^{-11}\,\mathrm{erg}\,\mathrm{s}^{-1}\,\mathrm{cm}^{-2}$
in order to remove any possible bias introduced by the upper limits or
large uncertainties at lower fluxes. Furthermore, we ignored data at
$F_\mathrm{PL}>3\times10^{-9}\,\mathrm{erg}\,\mathrm{s}^{-1}\,\mathrm{cm}^{-2}$
due to an insufficient description of the observed parameter evolutions at
these high fluxes regardless of the chosen model. We model and discuss
this discrepancy further below. We find that the photon index evolution,
$\Gamma(F_\mathrm{PL})$, is described best using a logarithmic function
($\rm log$: $\chi^2 = 97.0$; $\rm pow$: $\chi^2 = 183.1$; $\rm lin$:
$\chi^2 = 399.3$; all with 28 degrees of freedom). A power-law dependency
fits the evolution of the black-body flux, $F_\mathrm{BB}(F_\mathrm{PL})$,
well ($\rm pow$: $\chi^2 = 46.4$; $\rm lin$: $\chi^2 = 69.1$; $\rm log$:
$\chi^2 = 185.9$; all with 28 degrees of freedom).

Although the data follow these models well at first glance (see
Fig.~\ref{fig:parfluxcorr}) the goodnesses of the fits as given above
are not acceptable. A reduced $\chi^2$ near unity is, however,
required for a reasonable interpretation of the resulting model
parameters within their uncertainties.  The reason for the large
$\chi^2$-values are a few data points with very small uncertainties
compared to the complete dataset. These data points are, however, only
a few percent off from the models. We have tried to fit these
differences by introducing calibration constants similar to those
during the spectral analysis (see Sect.~\ref{sec:spec}). This approach
failed due to the low number of data points for \nustar and \suzaku
compared to \rxte and due to very similar spectral parameters among
these missions at high luminosities of \gro. To take into account the
systematic effect of these offsets, we have added a systematic
uncertainty of 0.03 in $\Gamma$ (corresponding to 2--7\% relative
uncertainty) and a relative uncertainty of 3.5\% in both flux
parameters, $F_\mathrm{PL}$ and $F_\mathrm{BB}$. These additional
uncertainties are consistent with known energy and flux
cross-calibration uncertainties between different X-ray missions and
their instruments (see, e.g., \citealt{kirsch2005a},
\citealt{tsujimoto2011a}, or \citealt{madsen2017a}). We note that we
cannot exclude source variability on a few percent level as a reason
for the large $\chi^2$-values besides calibration uncertainties. This
does not, however, affect any of the following conclusions given the
overall change in $\Gamma$ and $F_\mathrm{BB}$ by a factor of 2--3 and
by a few orders of magnitude, respectively.

When including all available data over the full flux range (with the
exception of epoch~3), we find that the recent observations (when the
source was at very high luminosities), are not consistent with a
logarithmic function for $\Gamma(F_\mathrm{PL})$ and, especially, a
power-law function for $F_\mathrm{BB}(F_\mathrm{PL})$ with a single
slope. This is similar to what we found for lower fluxes in
\citetalias{kuehnel2014a}. For these three epochs (1a, 1b, and 4) we
measure the same parameter values within their uncertainties, despite
taken at very different fluxes. This behavior can be fitted much
better by a flattening of the correlation towards higher fluxes,
$F_\mathrm{brk}$,
\begin{equation}\label{eq:corrfunsbrk2}
  {\rm high\hbox{-}brkn\!}: X^\prime(F_\mathrm{PL}) = \begin{cases}
     X(F_\mathrm{PL}), & {\rm for~} F_\mathrm{PL} \le F_\mathrm{brk,hi}, \\
     X(F_\mathrm{brk,hi}), & {\rm for~} F_\mathrm{PL} > F_\mathrm{brk,hi}.
  \end{cases}
\end{equation}
Interestingly, the best-fit break fluxes for $\Gamma(F_\mathrm{PL})$ and
$F_\mathrm{BB}(F_\mathrm{PL})$ are the same within their uncertainties.
Thus, we tied the break fluxes, $F_\mathrm{brk,hi}$, for both parameter
evolutions together. The observed parameter correlations can be
described even better if we allow for an additional break of the photon
index, $\Gamma$, to a constant at lower fluxes, $F_\mathrm{brk,lo}$,
which we had noticed already in \citetalias{kuehnel2014a},
\begin{equation}\label{eq:corrfunsbrk1}
  {\rm low\hbox{-}brkn\!}: X^\prime(F_\mathrm{PL}) = \begin{cases}
     X(F_\mathrm{brk,lo}), & {\rm for~} F_\mathrm{PL} < F_\mathrm{brk,lo}, \\
     X(F_\mathrm{PL}), & {\rm for~} F_\mathrm{PL} \ge F_\mathrm{brk,lo}.
  \end{cases} \\
\end{equation}
This model for the spectral evolution of \gro including two breaks at
different luminosity levels provides a good description of the
data. The corresponding fit parameters are listed in
Table~\ref{tab:corrmodels}.

Since we interpret the artificial breaks in the model function as
transitions between different accretion regimes, the statistical
significance of their detection is crucial for our conclusion. The
least prejudiced way to derive the significance of these model
components is a Monte Carlo approach. Therefore, we simulated
$1.77\times10^6$ data sets for both, $\Gamma(F_\mathrm{PL})$ and
$F_\mathrm{BB}(F_\mathrm{PL})$, based on our best-fit model
\emph{without} any breaks. A Gaussian randomization of the individual
data points with their respective uncertainty have been applied for
each simulated dataset. The resulting data sets were then fitted both
with the model with and without the two breaks. If the $\chi^2$
difference of those two fits was larger than the one obtained from
fitting our measured data, we recorded this as a false-positive
detection. The fraction of false-positive detections in the complete
simulation directly translates to the significance of the modeled
break. We find a significance of the high-luminosity break in
$F_\mathrm{BB}$ of ${\ge}5\sigma$ as a lower limit. The high- and
low-luminosity breaks of the photon index, $\Gamma$, are significant
at the $2.35\sigma$ and $3.73\sigma$ level, respectively. We thus
conclude that the high-luminosity break in $F_\mathrm{BB}$ and the
low-luminosity break in $\Gamma$ are most likely real, while the
high-luminosity break of $\Gamma$ is only moderately significant. Note
that the chosen additional uncertainties as described above move these
significances to the conservative side. For instance, the significane
of the low-luminosity break in $\Gamma$ increases to $4.12$ once no
systematic uncertainty is added and only \rxte data are taken into
account (this is the only data set relevant at this luminosity level,
see Fig.~\ref{fig:parfluxcorr}).

We stress that other models for the spectral evolution of \gro as a
function of luminosity might provide an equivalent description of the
data. However, regardless of the chosen model or approach (e.g.,
investigating hardness ratios) we significantly detect changes near
$F_\mathrm{brk,lo}$ and $F_\mathrm{brk,hi}$. We do not claim that our
(phenomenological) model or the way in handling systematic
uncertainties is generally valid. Thus, the spectral shape of \gro for
any flux as predicted by our model should be taken with care.

\subsection{Accretion Regimes in \gro}\label{sec:regimes}

The detailed analysis of the evolution of the photon index and black
body flux presented in Sect.~\ref{sec:discuss:parevol} revealed two
flux levels, where a change in the evolution is happening (see
Fig.~\ref{fig:parfluxcorr}). At fluxes below
$F_\mathrm{brk,lo} \sim 10^{-10}$\,erg\,s$^{-1}$\,cm$^{-2}$
(between 15--50\,keV) the photon index seems to be independent of
flux and stays constant. For higher fluxes, the photon index starts to
harden, while the black body flux increases. Once the flux exceeds
$F_\mathrm{brk,hi} \sim 3\times 10^{-9}$\,erg\,s$^{-1}$\,cm$^{-2}$ a
saturation effect is observed, where the photon index is no longer
hardening and the black body does not increase in flux further. In
addition, during the giant 2012 November outburst (epoch~1a and
1b), when \gro reached its highest known flux so far\footnote{The
physical flux of \gro within 20--50\,keV  during its discovery
outburst in 1993 was $2.2\times 10^{-9}$\,erg\,s$^{-1}$\,cm$^{-2}$
\citep{shrader1999a}.}, another soft component below 3\,keV shows up
in the \suzaku-spectrum (the second black body, see
Sect.~\ref{sec:spec:suzaku}). This feature is not detected in any
other observation at lower fluxes. We interpret these facts as
observational evidence for three different accretion regimes in \gro.

\begin{figure}
  \includegraphics[width=\columnwidth]{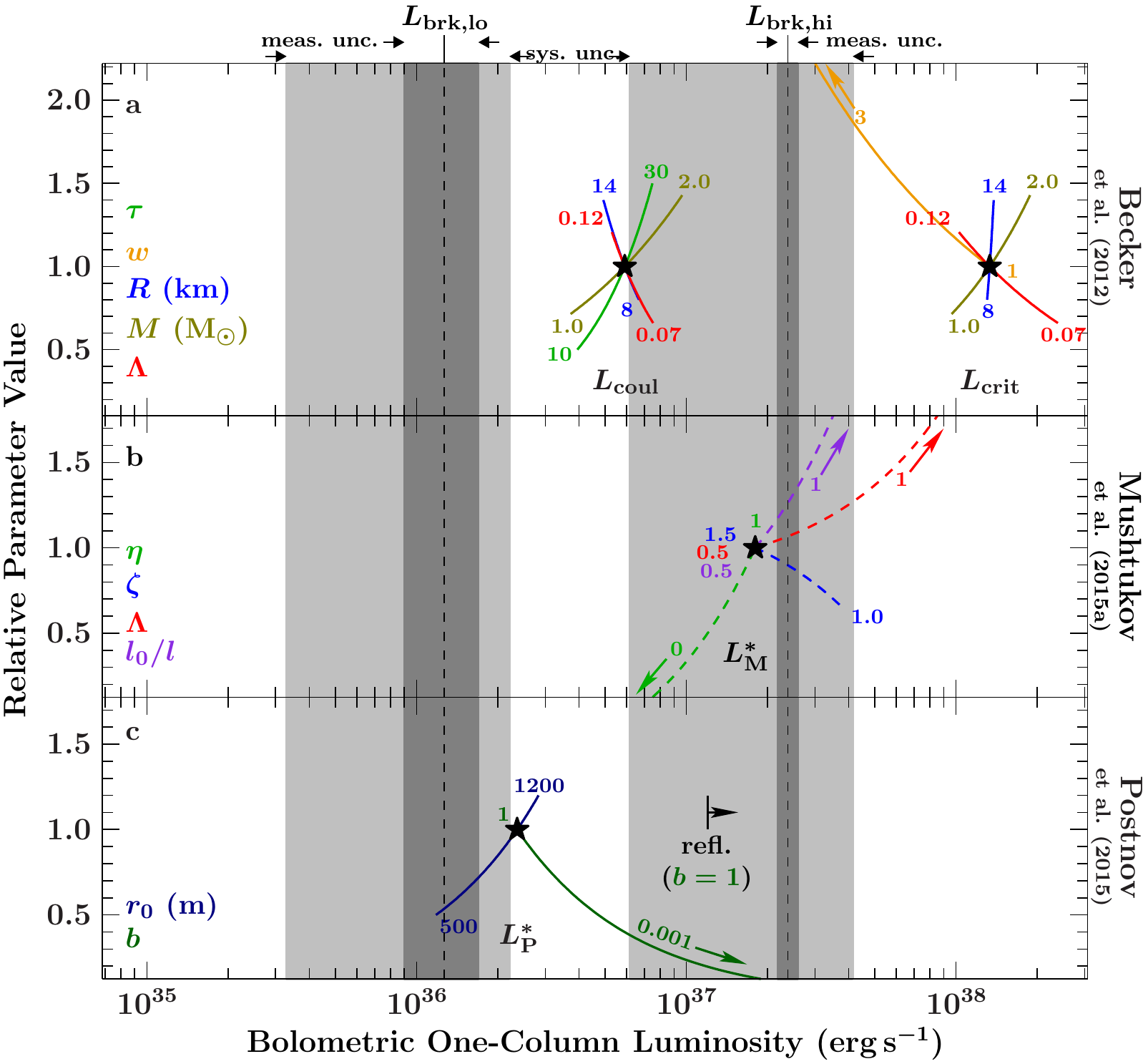}
  \caption{Comparison of the derived luminosities at the breaks of
  the spectral evolution ($L_\mathrm{brk,lo}$ and
  $L_\mathrm{brk,hi}$, vertical dashed lines) with $L_\mathrm{coul}$
  and $L_\mathrm{crit}$ of \citet[][Eqs.~\ref{eq:lcoul} \&
  \ref{eq:lcrit}; panel \textbf{a}]{becker2012a}, $L^\ast_\mathrm{M}$ of
  \citet[][Eq.~\ref{eq:lastcrit}; \textbf{b}]{mushtukov2015a}, and
  $L^\ast_\mathrm{P}$ of \citet[][Eq.~\ref{eq:last}; \textbf{c}]{postnov2015b}.
  The dark gray bands correspond to the 90\% statistical
  uncertainties of the measured break fluxes (meas.\ unc). The
  larger widths in light gray are the uncertainties when deriving
  luminosities from the observed fluxes as described in the text
  (sys.\ unc.). The black stars are the theoretically
  expected values for typical neutron star parameters and
  $E_\mathrm{cyc}$ as found by \nustar and corrected for the
  gravitational redshift. The colored lines represent changes in the
  neutron star parameters with their relative value along the y-axis.
  The minimum and maximum absolute values are marked at the line endings.
  The black arrow in panel c marks the luminosity above which
  \citet{postnov2015b} expect reflected radiation from the neutron
  star's surface to contribute significantly. The dependencies of the
  parameters in panel b (dashed lines) are assumed to be 
  linear (see text for details).}\label{fig:comparetheory}
\end{figure}

In order to connect these different accretion regimes of \gro with
the physics of the accretion process, we have compared our findings with
recent theoretical investigations by \citet{becker2012a},
\citet{postnov2015b}, and \citet{mushtukov2015a}. These authors estimate
the so-called critical luminosity, which can be interpreted as a
transition between two accretion regimes: above this luminosity the
radiation pressure inside the column dominates the deceleration of the
infalling plasma, while Coloumb interactions dominate at lower
luminosities. In addition, \citet{becker2012a} expect these Coloumb
interactions to be unable to stop the material to rest for even lower
luminosities. This so-called Coloumb braking luminosity would mark
another transition between accretion regimes. The theoretical
assumptions and calculations of these three works result, however, in
differences by orders of magnitude. More details and the actual
equations for the transition luminosities are given in
Appendix~\ref{app:lcrit}.

Figure~\ref{fig:comparetheory} compares the luminosities of \gro at
the observed breaks in the spectral parameter evolution,
$L_\mathrm{brk,lo}$ and $L_\mathrm{brk,hi}$, with the theoretically
expected luminosities at the transition between different accretion
regimes after \citet{becker2012a}, \citet{postnov2015b}, and
\citet{mushtukov2015a} (Eqs.~\ref{eq:lcoul}--\ref{eq:lastcrit}).
To derive the bolometric luminosities in the neutron star's
rest-frame, $L$, from the observed fluxes, $F_\mathrm{brk,lo}$ and
$F_\mathrm{brk,hi}$, we need to
\begin{enumerate}[a)]
  \item extrapolate the model to the full electromagnetic
  spectrum in order to calculate the unabsorbed bolometric flux,
  $F^\star$,
  \item correct for the gravitational redshift, $z$,
  \item take into account a factor, $g$, describing the emission
  geometry,
  \item and, correct for the distance, $d$, to the source.
\end{enumerate}
In summary, 
\begin{equation}\label{eq:flux2lum}
  L = g\,d^2 (1+z)^2 F^\star(F_\mathrm{PL}).
\end{equation}
We describe the details of the corrections a) to d) in the following
and discuss their uncertainties.

\noindent a) The unabsorbed bolometric flux, $F^\star$, is a
non-linear function of the power-law flux, $F_\mathrm{PL}$. We have
used our spectral evolution model (see Table~\ref{tab:corrmodels}) in
order to extrapolate the observed spectral shape to the 0.01 to
100\,keV energy range. From this extended spectrum we determine the
bolometric flux. The uncertainty of the extrapolation is caused by the
unknown spectral shape outside of the observed energy band. Here, we
assume a thermal Comptonization spectrum\footnote{\texttt{compTT}
  \citep{titarchuk1994a}: seed photons of $kT = 1.883$\,keV, plasma
  temperature of $E_\mathrm{fold} = 15.9$\,keV, plasma optical depth
  of 20 \citep{becker2012a}, and normalized to match the source's
  spectrum at 1\,keV.} below 1\,keV, i.e., below the \swift-XRT and
\suzaku-XIS sensitivities. Furthermore, the uncertainty in the
estimated bolometric flux takes the uncertainty of the measured
absorption column density, $N_\mathrm{H}$, into account. It is
unlikely that the observed X-ray absorption mimics an actual
Comptonization turn-over since we do not observe a change in
$N_\mathrm{H}$ with time or flux in \gro
\citepalias{kuehnel2013a}. Due to the exponential rollover at higher
energies the spectral shape above ${\sim}80$\,keV (\mbox{\nustar} and
\suzaku-GSO) does not influence the bolometric flux significantly. We
found that the extrapolation of the energy band results in a
0.01--100\,keV flux uncertainty of 15\% when comparing thermal
Comptonization with the power-law spectrum used for spectral analysis
(see Sect.~\ref{sec:spec}).

\noindent b) The gravitational redshift, $1+z=(1-2 G M / R
c^2)^{-1/2}$ with the mass, $M$, and radius, $R$, of the neutron star
results in an observed luminosity of $L_\mathrm{obs} = L / (1+z)^2$
compared to the intrinsic one \citep{thorne1977a}. For a typical
neutron star with $M=1.4\,\msol$ and $R=12$\,km, we find $z\approx
0.235$. Assuming $\Delta M = 0.5\,\msol$ and $\Delta R = 6$\,km we
find an uncertainty of $\Delta z= 0.1$ in $z$, which corresponds to a
17\% uncertainty in the derived intrinsic luminosity.

\noindent c) The emission geometry of \gro is unknown and, thus, we
assume isotropic emission of each pole into its hemisphere, i.e.,
$g=2\pi$ (the theoretical expectations apply to \textit{one} pole
only). According to \citet{martineznunez2016a}, who estimate the
effects of light bending \citep[e.g.,][]{kraus2001a}, the real
luminosity for a typical fan beam accretion geometry can differ by
${\sim}25\%$ relative to the derived luminosity assuming isotropic
emission. Due to the higher accretion rate in BeXRBs compared to low
mass X-ray binaries (LMXBs), which usually emit via a pencil beam
geometry \citep[see, e.g.,][]{basko1975a}, the fan beam is a justified
geometry for \gro. Thus, we assume a systematic uncertainty of 25\%
due to the unknown emission geometry.

\noindent d) The distance to the source was determined to
$d=5.8(5)$\,kpc by \citet{riquelme2012a}, which propagates to a
further 17\% uncertainty in the luminosity.

Using Eq.~\ref{eq:flux2lum} we converted the measured fluxes where we
observed changes in the spectral behavior of \gro, $F_\mathrm{brk,lo}$
and $F_\mathrm{brk,hi}$, into the luminosities shown in
Fig.~\ref{fig:comparetheory}, $L_\mathrm{brk,lo}$ and
$L_\mathrm{brk,hi}$ (vertical dashed lines). Besides the corresponding
measurement uncertainties of these luminosities (dark gray bands) we
added a systematic uncertainty of 74\% to the data (light gray bands)
which corresponds to the sum of the additional sources of uncertainty,
a--d, as described above\footnote{The uncertainties of the
contributions a--d are mainly due to systematics caused by, e.g.,
extrapolation or theoretical assumptions. Thus, we chose to add their
corresponding uncertainties directly instead of adding them in
quadrature.}. In order to calculate the theoretical luminosities after
Eqs.~\ref{eq:lcoul}--\ref{eq:lastcrit} (black stars), we assumed
typical neutron star parameters (see
Eqs.~\ref{eq:lcoul}--\ref{eq:lastcrit} and the respective references)
with the exception of the surface cyclotron line energy, $E_\star$,
which we assumed to be the redshift corrected cyclotron line energy,
$E_\star = (1+z) E_\mathrm{cyc} = 90$\,keV with the mean value of
$E_\mathrm{cyc}$ as listed in Tab.~\ref{tab:bestfits}). The
dependencies of the different luminosities on the neutron star
parameters are shown by the colored lines. Because
\citet{mushtukov2015a} solved Eq.~\ref{eq:lastcrit} numerically they
presented the dependencies of their critical luminosity,
$L^\ast_\mathrm{M}$, on the neutron star parameters in their Fig.~5
instead of providing an analytical equation. From this Figure, we have
extracted the values of the critical luminosity assuming the same
redshift corrected cyclotron line energy as above. In order to include
the dependency of the critical luminosity, $L^\ast_\mathrm{M}$
(Eq.~\ref{eq:lastcrit}), on neutron star parameters in
Fig.~\ref{fig:comparetheory}, we assumed a linear dependency between
the extracted values. By drawing the resulting dependencies as dashed
lines (panel b) we stress that this is for illustrating the basic
dependence, i.e., its sign only.

The facts from Fig.~\ref{fig:parfluxcorr}, which shows the spectral
parameter evolution of \gro and Fig.~\ref{fig:comparetheory}, which
compares the changes in this parameter evolution at
$L_\mathrm{brk,lo}$ and $L_\mathrm{brk,hi}$ with recent theoretical
investigations, can be summarized as follows.
\begin{itemize}
  \item Between the observed luminosities $L_\mathrm{brk,lo}$ and
  $L_\mathrm{brk,hi}$ the spectrum of \gro is hardening with
  increasing luminosity. This is in line with theoretical
  expectations by \citet{postnov2015b} for sources in the
  subcritical accretion regime.
  \item If $L_\mathrm{brk,hi}$ can indeed be associated with the
  source exceeding the critical luminosity, then parameter
  combinations can be found such that each of the three theories
  summarized in Appendix~\ref{app:lcrit} predict the expected value for
  this luminosity correctly:
  \begin{itemize}
    \item $L_\mathrm{brk,hi} {\sim} L_\mathrm{crit}$ for
    $w > 1.9$ after
    \citet[Eq.~\ref{eq:lcrit}]{becker2012a}, i.e., the spectrum inside the
    column is a mixture between a Plank spectrum ($w=3$) and pure
    Comptonized bremsstrahlung-radiation ($w=1$).
    \item $L_\mathrm{brk,hi} {\sim} L^\ast_\mathrm{M}$ after
    \citet[Eq.~\ref{eq:lastcrit}]{mushtukov2015a} for canonical
    neutron star parameters.
    \item $L_\mathrm{brk,hi} {\sim} L^\ast_\mathrm{P}$ after
    \citet[Eq.~\ref{eq:last}]{postnov2015b} for $b \ll 1$, i.e., a
    hollow accretion column. Note that Eq.~\ref{eq:last} depends
    equally on $b$ and $\kappa_\perp/\kappa_\mathrm{T}$, i.e., the
    same effect is achieved for $\kappa_\perp/\kappa_\mathrm{T} \ll
    1$.
  \end{itemize}
  \item The change of the spectral evolution at $L_\mathrm{brk,lo}$
  is consistent with $L^\ast_\mathrm{P}$ after
  \citet[Eq.~\ref{eq:last}]{postnov2015b} for a filled accretion column
  with a radius of $r_0 \le 935$\,m. The corresponding
  luminosity around $10^{36}\,\mathrm{erg}\,\mathrm{s}^{-1}$ is,
  however, too low for the transition to supercritical accretion as
  expected by all these theories. Even \citet{postnov2015b} expect this
  transition to occur around $10^{37}\,\mathrm{erg}\,\mathrm{s}^{-1}$.
  \item The ratio between the Coulomb braking and the critical
  luminosity, $L_\mathrm{coul}/L_\mathrm{crit}$, after
  \citet[Eqs.~\ref{eq:lcoul} and \ref{eq:lcrit}]{becker2012a}
  matches almost perfectly the observed ratio
  $L_\mathrm{brk,lo}/L_\mathrm{brk,hi}$. The individual theoretical
  luminosities for canonical neutron star parameters are, however, a
  factor of ${\sim}6$ higher than compared to observed ones.
  \item The observed spectral change at high luminosities,
  $L_\mathrm{brk,hi}$, is also consistent with the saturation of the
  hardness as expected by \citet{postnov2015b} due to reflection
  from the neutron star's surface in case of a filled accretion
  column.
\end{itemize}
We have further investigated the dilemma that $L_\mathrm{brk,hi}$ is
consistent with reflection from a filled accretion column, but also
with $L^\ast_\mathrm{P}$ for a hollow accretion column after
\citet{postnov2015a}. From our best-fit parameter evolution (see
Table~\ref{tab:corrmodels}) we calculated the hardness ratio using the
same energy bands as \citet[][5--12\,keV over
1.3--3\,keV]{postnov2015a}. The resulting evolution of the hardness
ratio does not increase above a hardness of ${\sim}6$. This is in very
good agreement with the expected value for the hardness ratio in case
of a filled accretion column, whereas a hardness in the range of
10--16 is expected for a hollow accretion column \citep[see Figs.~6
and 7 of][]{postnov2015a}. Furthermore, the derived hardness saturates
above a (one-column) accretion rate of $1\times 10^{17}$\,g\,s$^{-1}$,
while the evolution of the hardness shown in Fig.~6 of
\citet{postnov2015a} suggests a saturation above
5--7$\times 10^{17}$\,g\,s$^{-1}$. This value scales with the height
of the filled accretion column, which itself anti-correlates with the
magnetic field strength at a given mass accretion rate (K.~Postnov,
priv.\ comm.). The required magnetic field strength to achieve a match
between the observed $L_\mathrm{brk,hi}$ and the expected saturation
of the hardness due to reflection is, however, far lower than compared
to \gro's CRSF. In summary, the match of $L^\ast_\mathrm{P}$ with
$L_\mathrm{brk,hi}$ in case of a hollow column is ruled out by the
observed hardness ratio. Once a filled column is assumed, the overall
hardness ratio matches the predicted value. The dependence of
$L^\ast_\mathrm{P}$ on the luminosity due to reflection at the neutron
star's surface requires, however, a much weaker magnetic field than
what is found for \gro.

We notice that the observed break at high luminosities,
$L_\mathrm{brk,hi}$, agrees well with the critical luminosity,
$L^\ast_\mathrm{M}$, as predicted by \citet{mushtukov2015a}, at least
for the case of \gro. The theories by \citet{postnov2015a} and
\citet{becker2012a} have difficulties explaining our observations
since unlikely parameter combinations are necessary in order to
achieve a match with their predicted critical luminosities,
$L^\ast_\mathrm{P}$ and $L_\mathrm{crit}$, respectively (see
Fig.~\ref{fig:comparetheory}). The large systematic uncertainties when
deriving luminosities from observational data as discussed above do
not, however, allow us to favor unequivocally one of the discussed
theories for the prediction of the critical luminosity. Especially,
the observed change of the spectral evolution of \gro at low
luminosities, $L_\mathrm{brk,lo}$, cannot be explained by either
theory. Furthermore, drawing general conclusions about these theories
is statistically questionable as only observational data of a single
source are used. In future work, a detailed study of the recently
claimed accretion regimes in 4U~1901+03 \citep{reig2016a} and
V~0332+53 \citep{doroshenko2016a} combined with the results by
\citet{reig2013a} for various sources could help clarify this
question. It should be noted, however, that thorough analysis methods
are required and systematic effects (caused by detector calibration
and due to the conversion to luminosities) have to be taken into
account as, e.g., discussed here. Nevertheless, the existence of
different accretion regimes, which are driven by the mass accretion
rate, seems to be confirmed.

In summary, however, the theories allow us to associate the following
physical accretion regimes with the observed changes in the spectral
evolution of \gro:
\begin{itemize}
  \item $L > L_\mathrm{brk,hi} \sim
  2{\times} 10^{37}\,\mathrm{erg}\,\mathrm{s}^{-1}$:
  Supercritital accretion, where radiation dominates the
  deceleration of the infalling plasma and stops the hardening of
  the X-ray spectrum.
  \item $10^{36}\,\mathrm{erg}\,\mathrm{s}^{-1} \sim
  L_\mathrm{brk,lo} < L < L_\mathrm{brk,hi}$:
  Subcritical accretion regime, where Comptonization effects scale
  with the mass accretion rate.
  \item $L < L_\mathrm{brk,lo}$: Very low subcritical regime, where
  physical effects depend only marginally on the mass accretion rate.
\end{itemize}
Further investigations of the source are required to confirm or
reject these conclusions. In order to proceed, however, working
self-consistent models should be applied to the spectra of \gro to
reveal the evolution of its \textit{physical parameters} with
luminosity. Additionally, high SNR observations of the source at
very low ($L < L_\mathrm{brk,lo}$) and extremely high luminosities
($L > L_\mathrm{brk,hi}$) are required to investigate the existence
of the different accretion regimes as proposed here. Finally, theory
predicts different flux dependencies of the CRSF parameters
depending on the accretion regime, which could be probed for \gro
with future hard X-ray missions with a sufficiently high effective
area around 100\,keV.

\begin{acknowledgement}
MK acknowledges support by the Bundesministerium f\"ur Wirtschaft und
Technologie under Deutsches Zentrum f\"ur Luft- und Raumfahrt grants
50OR1113 and 50OR1207. We appreciate the useful discussions with
K.~Postnov and M.~Gornostaev about accretion column physics. All figures
shown in this paper were produced using the \texttt{SLXfig} module,
developed by John E. Davis. We thank the \suzaku- and \nustar-teams for
accepting our proposals and performing the observations. Finally, we
thank the referee for her/his valuable comments, which helped improving
the quality of our paper.
\end{acknowledgement}

\appendix
\section{Critical Luminosity}\label{app:lcrit}

At the critical luminosity of an accreting neutron star the radiation
pressure generated at the base of the accretion column contributes
significantly to the deceleration of the infalling plasma
\citep{basko1975a}. In the past many attempts have been made to
derive this critical luminosity theoretically. Three of these theories
are summarized in the following.

The X-ray spectra of several accreting neutron stars exhibit so-called
cyclotron resonant scattering features (CRSFs; see, e.g.,
\citealt{caballero2012a} for a review). These absorption features arise
from transitions between the Landau levels of electrons in the accreted
plasma, which are quantized due to the strong magnetic field of these
neutron stars on the order of $10^{12}$\,G. The observed cyclotron line
energy, $E_\mathrm{cyc}$, is found to show a positive or negative
correlation with the mass accretion rate, $\dot{M}$ (see, e.g.,
\citealt{caballero2012a} and \citealt{becker2012a}, and references
therein).

In order to explain theoretically the different types of behavior of
CRSF energy with luminosity, \citet{becker2012a} investigated the
characteristic height of the X-ray emission region in the accretion
column as a function of the mass accretion rate. These authors proposed
different $\dot{M}$ regimes in which the height of the shock positively
correlates (negative $E_\mathrm{cyc}$ correlation) or negatively
correlates (positive $E_\mathrm{cyc}$ correlation) with $\dot{M}$. The
luminosities at which the transitions between these regimes occur, are
known as $L_\mathrm{coul}$ and $L_\mathrm{crit}$, and are given by
\citep[Eqs.~59 and 55 of][]{becker2012a}
\begin{multline}\label{eq:lcoul}
  L_\mathrm{coul} = 1.23 \times 10^{37}\,\mathrm{erg}\,\mathrm{s}^{-1}
  \left(\frac{\Lambda}{0.1}\right)^{-7/12}
  \left(\frac{\tau_\star}{20}\right)^{7/12}\\
   \left(\frac{M_\star}{1.4\,\msol}\right)^{11/8}
  \left(\frac{R_\star}{10\,\mathrm{km}}\right)^{-13/24}
  \left(\frac{E_\star}{10\,\mathrm{keV}}\right)^{-1/3}
\end{multline}
and
\begin{multline}\label{eq:lcrit}
  L_\mathrm{crit} = 1.28 \times 10^{37}\,\mathrm{erg}\,\mathrm{s}^{-1}
  \left(\frac{\Lambda}{0.1}\right)^{-7/5}
  w^{-28/15}
  \left(\frac{M_\star}{1.4\,\msol}\right)^{29/30}\\
   \left(\frac{R_\star}{10\,\mathrm{km}}\right)^{1/10}
  \left(\frac{E_\star}{10\,\mathrm{keV}}\right)^{16/15}.
\end{multline}
Here, $\Lambda$ is a parameter describing the accretion geometry
($\Lambda=1$ for spherical accretion from a wind and $\Lambda<1$ for
accretion from a disk, see also \citealt{lamb1973a}), $\tau_\star$ is
the Thomson optical depth, $M_\star$ and $R_\star$ are the mass and
radius of the neutron star, respectively, $E_\star$ is the CRSF energy
related to the surface magnetic field, and $w$ a parameter describing
the spectral shape inside the column ($w=1$ for a Bremsstrahlung
spectrum and $w=3$ for a Planck spectrum). In the model of
\citet{becker2012a}, the in-falling matter is decelerated by passing
through a radiation dominated shock. Above the so-called critical
luminosity, $L_\mathrm{crit}$ (in the supercritical accretion regime),
the radiation pressure alone is able to stop the matter above the
neutron star's surface. Below this luminosity (in the subcritical
regime), the radiation dominated shock still exists, but the final
deceleration occurs via Coulomb braking within the accretion flow. At
very low luminosities, below a characteristic luminosity,
$L_\mathrm{coul}$, Coulomb interactions are no longer sufficient to stop
the matter. The detailed mechanism to decelerate the matter to rest is
not yet clear \citep[see., e.g.][]{fuerst2014a}.

Alternatively, \citet{mushtukov2015a} calculate the critical
luminosity, i.e., where the supercritical accretion sets in taking
resonant scattering and photon polarization into account for the
first time. For a circular hotspot on the neutron star's surface,
this luminosity is given by \citep[][Eq.~7 in
\citealt{mushtukov2015a}]{basko1975a,basko1976a}
\begin{equation}\label{eq:lastcrit}
  L^\ast_\mathrm{M} \approx 3.7 \times 10^{36}
  \left(\frac{\kappa_\mathrm{T}}{\kappa_\mathrm{eff}}\right)
  \left(\frac{d}{10^5\,\mathrm{cm}}\right)
  \left(\frac{R_\star}{10^6\,\mathrm{cm}}\right)^{-1}
  \left(\frac{M_\star}{1\,\msol}\right)\,\mathrm{erg}\,\mathrm{s}^{-1},
\end{equation}
where $\kappa_\mathrm{eff}$ is the effective scattering cross-section
and $d$ is the diameter of the hot spot. The key issue here is the
calculation of the effective scattering cross-section,
$\kappa_\mathrm{eff}$, which is solved by \citet{mushtukov2015a}
numerically assuming specific accretion column geometries for wind- and
disk-accretion, a linear velocity profile, and black body seed photons.
Although these authors assumed a radiation dominated shock to exist for
luminosities above the critical luminosity, $L^\ast_\mathrm{M}$, they
argue that the value of $L^\ast_\mathrm{M}$ is mainly determined by the
effective cross-section, $\kappa_\mathrm{eff}$, due to resonant
scattering of electrons. In contrast to \citet{becker2012a}, they expect
the X-ray emission region to settle down on the neutron star's surface
for luminosities below the critical luminosity, $L^\ast_\mathrm{M}$. In
this subcritical regime, \citet{mushtukov2015b} do not expect a
radiation dominated shock to be formed. Instead, they explain the
positive correlation of $E_\mathrm{cyc}$ with $\dot{M}$ by the (still)
relativistic velocity of the in-falling plasma near the surface, which
results in a Doppler shift of the CRSF energy.

Another recent theoretical investigation by \citet{postnov2015b}
calculates the X-ray spectrum in the supercritical regime using the
radiation diffusion approximation and assuming the emission emerging
from the walls of the accretion column (known as the fan beam geometry).
In this scenario the accretion column is assumed to be optically thick
and the infalling matter is decelerated by a radiative shock similar to
\citet{becker2012a}. The minimum luminosity, $L^\ast_\mathrm{P}$, at
which an optically thick accretion column appears, is given by
\citep[Eq.~5 of][]{postnov2015b}
\begin{equation}\label{eq:last}
  L^\ast_\mathrm{P} \approx 2.36 \times 10^{36}\,\mathrm{erg}\,\mathrm{s}^{-1}
  \left(\frac{r_0}{10^5\,\mathrm{cm}}\right)
  \left(\frac{b \kappa_\perp}{\kappa_T}\right)^{-1},
\end{equation}
with the radius $r_0$ of the accretion column, the cross-sections
$\kappa_\perp$ and $\kappa_T$, and the thickness of the column walls, $0
< b \le 1$, relative to its radius ($b=1$ corresponds to a filled
accretion column, while $b \ll 1$ is a hollow column). In principle,
their derived luminosity, $L^\ast_\mathrm{P}$, describes the same
physical condition as $L_\mathrm{crit}$ after \citet{becker2012a} and
$L^\ast_\mathrm{M}$ after \citet{mushtukov2015a}. For luminosities $L <
L^\ast_\mathrm{P}$, i.e., in the subcritical regime,
\citet{postnov2015b} showed that the observed X-ray spectrum hardens
with increasing mass accretion rate, $\dot{M}$, due to an increase in
the Comptonization parameter. Finally, at very high luminosities around
(3--7)$ \times 10^{37}\,\mathrm{erg}\,\mathrm{s}^{-1}$ they observe a
saturation in the hardness in some accreting pulsars using data from the
All Sky Monitor (ASM) onboard the \textit{Rossi X-ray Timing Explorer}
\citep[\rxte][]{bradt1993a}. They are able to explain this behavior by
including reflected radiation from the neutron star's surface, which is
illuminated by the Doppler-boosted radiation of the column walls, while
the spectrum continues to harden for increasing luminosities. Thus, in
the model of \citet{postnov2015b}, the observed saturation is a purely
geometric effect.

\end{document}